\documentclass[useAMS,usenatbib]{mn2e}
\usepackage{amsmath}
\usepackage{amsfonts}%
\usepackage{amssymb}
\usepackage{graphicx, natbib, wasysym, subfiles, aas_macros}

\newcommand{\parti}[2]{\partial_{#2} #1}
\renewcommand{\vec}[1]{\ensuremath{\boldsymbol{#1}}} 
\DeclareMathAlphabet{\textbfsf}{\encodingdefault}{\sfdefault}{bx}{sl}
\newcommand{\tensor}[1]{\ensuremath{\textbfsf{#1}}} 

\usepackage{ulem}

\def\parti#1#2{\frac{\partial #1}{\partial #2}}

\def\r{{\bf r}}
\usepackage{color}

\title[Eccentricity of Massive Planets ]{Eccentricity and Inclination of Massive Planets  Inside Low-density Cavities: Results of 3D Simulations}

\author[Romanova et al.]{\parbox{\textwidth}{M. M.~Romanova$^{1,2}$\thanks{E-mail of
corresponding author: \texttt{romanova@astro.cornell.edu}},
 A. V.~Koldoba$^{3}$, G. V.~Ustyugova$^{4}$, C. Espaillat$^{5}$, R. V. E.~Lovelace$^{1,2}$}
\vspace{0.4cm}\\
\parbox{\textwidth}{ 
$^{1}$Department of Astronomy, Cornell University, Ithaca, NY 14853-6801\\
$^{2}$Carl Sagan Institute, Cornell University, Ithaca, NY 14853-6801\\
$^{3}$Moscow Institute of Physics and Technology, Dolgoprudny, Moscow Region, 141700, Russia \\
$^{4}$Keldysh Institute for Applied Mathematics, Moscow, 125047,
Russia \\
$^{5}$Department of Astronomy, Boston University,
725 Commonwealth Avenue
Boston, MA 02215  }}
\date{\today}

\pagerange{\pageref{firstpage}--\pageref{lastpage}} \pubyear{2002}

\begin{document}
\label{firstpage}

\maketitle

\begin{abstract}

\noindent We study the evolution of eccentricity  and inclination of massive planets in low-density cavities of  protoplanetary discs using three-dimensional (3D) simulations. When the planet's orbit is \textit{aligned} with the equatorial plane of the disc, the eccentricity increases to high values of 0.7-0.9 due to the resonant interaction with the inner parts of the disc. 
For planets on \textit{inclined} orbits, the eccentricity increases due to the Kozai-Lidov mechanism, where the disc acts as an external massive body, which perturbs the planet's orbit.  
At small inclination angles, $\lesssim 30^\circ$, the resonant interaction with the inner disc strongly  contributes to the eccentricity growth, while at larger angles, eccentricity growth is mainly due to the Kozai-Lidov mechanism.  
We conclude that planets inside low-density cavities tend to acquire high eccentricity if favorable conditions give sufficient time
for growth. The final value of the planet's eccentricity after the disc 
dispersal,
depends on the planet's mass and properties of the cavity and protoplanetary disc.

\end{abstract}


\section{Introduction}

Many exoplanets have high eccentricities of their orbits.
Giant planets 
have eccentricities covering the whole range from zero to near unity
(e.g., \citealt{MarcyEtAl2005,
KaneEtAl2012,SagearBallard2023}). The phenomenon of non-zero eccentricities has yet to be understood. 

Eccentricity and inclination may grow due to the gravitational interaction between planets
 (e.g., \citealt{RasioFord1996,LinIda1997,PapaloizouTerquem2001,ChatterjeeEtAl2008,JuricTremaine2008,MustillEtAl2017,AndersonEtAl2020,
LiEtAl2021}) or due to secular perturbations from exterior stellar or planetary companions due to Kozai-Lidov mechanism  \citep{Kozai1962,Lidov1962,HolmanEtAl1997,TakedaRasio2005,FabryckyTremaine2007, AndersonEtAl2016,AndersonLai2017}.
On the other hand, they may vary due to  
the interaction of a planet with an accretion disc (e.g.,  \citealt{GoldreichTremaine1979}).

A planet in the low-density cavity 
 interacts with the inner disc by Eccentric Lindblad Resonances (ELRs), and the eccentricity increases   (e.g., \citealt{GoldreichTremaine1979,GoldreichTremaine1980,ArtymowiczEtAl1991,GoldreichSari2003,OgilvieLubow2003,TeyssandierOgilvie2016}). 
A low-density cavity may be supported by various  physical  mechanisms, e.g.,  by the magnetosphere of the star  (e.g.,  \citealt{Konigl1991,Hartmann2000,RomanovaLovelace2006,RomanovaOwocki2015}), magnetic wind from the star (e.g., \citealt{LovelaceEtAl2008,SchnepfEtAl2015,Bai2016,WangGoodman2017,ElbakyanEtAl2022}), or
evaporation of the inner disc due to UV radiation (e.g., \citealt{DullemondEtAl2007}).

A number of two-dimensional (2D) numerical simulations have been performed that show that eccentricity can increase due to the disc-planet resonant interaction
 (e.g., \citealt{PapaloizouEtAl2001,dAngeloEtAl2006,KleyDirksen2006,RiceEtAl2008,BitschEtAl2013a,DunhillEtAl2013,RagusaEtAl2018,DebrasEtAl2021,BaruteauEtAl2021}). In many simulations, only a small value of eccentricity has been obtained, $e\sim 0.1- 0.25$ (e.g., \citealt{PapaloizouEtAl2001,dAngeloEtAl2006,KleyDirksen2006}). Larger eccentricity $e\approx 0.4$  has been observed by 
 \citet{DebrasEtAl2021} who  were able to keep the disc-cavity boundary at the same location by setting a high/low viscosity in the cavity/disc, 
and by \citet{RiceEtAl2008} 
who fixed the disc-cavity boundary and calculated the gas evolution only in the disc. The eccentricity of the planet may also increase or decrease due to the exchange of eccentricity between the planet and the disc
  (e.g.,  \citealt{TeyssandierOgilvie2016, RagusaEtAl2018,LiLai2023}).

In our earlier work, we used 2D simulations to investigate the eccentricity growth of massive planets located in the cavity with the fixed disc-cavity boundary (\citealt{RomanovaEtAl2023}, hereafter R23). We examined a wide range of parameters and observed that the eccentricity typically increases to high values of  $\sim 0.6-0.8$. 
We  investigated different resonances responsible for eccentricity growth and derived the dependence of the eccentricity growth on various parameters.  
Simulations confirmed the theoretically-predicted result that the eccentricity growth rate is proportional to the  density (characteristic mass)
 of the disc. Therefore, simulations are scalable and can be performed for denser discs during a shorter simulation time.    
 These findings opened a path for more realistic, 3D simulations of the disc-planet interaction.  

This paper shows the results of our global 3D simulations of the disc-planet interaction. We did not fix the disc-cavity boundary and calculate the gas flow inside and outside the cavity. The low-density cavity has been supported by the equilibrium initial conditions and a low disc viscosity. Simulations confirmed that the eccentricity may increase to very high values of  $e\sim 0.7-0.9$. We investigate the dependence of the growth rate of eccentricity on different factors, such as the planet's mass and the density, viscosity, and mass of the planet. In models with more massive planets and denser discs, the torques are higher, and eccentricity increases more rapidly. In models with lower grid resolution and higher viscosity, the resonances are not well resolved, and the eccentricity growth is slower.

We also investigate the evolution of eccentricity and inclination of planets in \textit{inclined} orbits.  
\citet{TerquemAjmia2010} have shown that the orbit of an inner planet can be perturbed by a remote massive disc due to the Kozai-Lidov mechanism
 similar to that when perturber is  a distant massive planet or a star. They supported their theoretical findings with N-body numerical simulations (see also \citealt{TeyssandierEtAl2013}). In their simulations, the external disc has been fixed. 
In our work, we calculate the general 3D hydrodynamic model where the density distribution in the disc changes due to the disc-planet interaction. 
We observed that the eccentricity increases and oscillates due to the Kozai-Lidov effect.  
We investigated the dependence of the eccentricity growth on the inclination of the orbit and other parameters. 

The plan of the paper is as follows. In Sec. \ref{sec:model}, we describe the problem setup and our numerical model.  In Sec. \ref{sec:aligned} and \ref{sec:inclined}, we show the results of simulations in cases of aligned and inclined orbits of the planet. 
We  
conclude in Sec. \ref{sec:conclusions}.

\section{Problem setup and numerical model}
\label{sec:model}

We place a star of mass $M_*$ in the center of the coordinate system.
We use a 3D grid in cylindrical coordinates $(r, \phi, z)$. 
We place a low-density cavity at radii 
$R_{\rm in}<r<r_{\rm cav}$, and high-density disc at radii  $r_{\rm cav}<r<R_{\rm o}$,  where $R_{\rm in}$ and $R_{\rm o}$ 
are the inner and outer boundaries of the simulation region.  

We solve a problem in dimensionless form. We measure distances in units of $r_0=r_{\rm cav}$.   The
inner and outer boundaries are 
 $R_{\rm in}=0.3$ and  $R_{\rm o}=18$.
The reference mass is the mass of the star, $M_0 = M_*=M_\odot$, where we take a Solar mass star as a base.
The reference velocity is given by $v_0 = \sqrt{GM_0/r_0}$. The time is measured 
in Keplerian periods of rotation at $r=r_0$: $P_0 = 2
\pi r_0/v_0$. The reference density is $\rho_0 = M_0 / r_0^3$ and
the reference surface density is $\Sigma_0 = M_0 / r_0^2$. The
reference pressure is $p_0 = \rho_0 v_0^2$.
We also determine the dimensionless mass of the inner disc, $q_d$, such that the dimensional characteristic mass of the inner disc is 
$M_{\rm d0}=q_d M_0$, where $M_{\rm d0}= \Sigma_{\rm d0} r_0^2$ is characteristic mass of the inner disc\footnote{Note that many authors use the total mass of the disc in their definition of $q_d$ (e.g., \citealt{TeyssandierOgilvie2016,RagusaEtAl2018}).} and , where $\Sigma_{\rm d0} =q_d \Sigma_0$ is the characteristic surface density, and therefore $q_d\equiv\widetilde{\Sigma}_d=\Sigma_{\rm d0}/\Sigma_0$ is also a dimensionless surface density of the disc at the reference point $r=r_0$. We drop tilde and hereafter use the dimensionless surface density $\Sigma_d$ as a parameter that characterizes typical mass of the inner disc. In current simulations, we take $\Sigma_d=3\times 10^{-2}$. For practical applications, we  use the more realistic value of $\Sigma_d=10^{-4}$ (see Tab. \ref{tab:units}).

 We place a planet of mass $m_p=q_p M_0=5\times 10^{-3}M_0=5 M_{\rm Jup}$  inside the cavity at the orbit with the semi-major axis $a_0=0.6$.
We take masses   $m_p=3 M_{\rm Jup}$ and  $m_p=10 M_{\rm Jup}$ in test simulation runs. We take an orbit with a small initial eccentricity $e=0.02$. It helps
to decrease the eccentricity damping by the 1st order corotation torque (e.g., \citealt{GoldreichSari2003,OgilvieLubow2003}).
A planet is placed either in the equatorial plane of the disc at zero inclination angle, $i_0=0^\circ$, or in inclined orbit with inclination angles from $i_0=5^\circ$ up to $i_0=75^\circ$.

\begin{table*}
\begin{tabular}[]{ cc | ccc }
\hline \hline \multicolumn{2}{c}{Reference Unit} &
\multicolumn{3}{c}{Reference Values}    \\ \hline
Reference distance  $r_0=r_{\rm cav}$   & $r_0$ [AU]            & 0.1                              & 1.0                              & 10                               \\
Reference velocity                    & $v_0$ [km s$^{-1}$]           & 94.3                             & 29.8                            & 9.4                                \\
Reference period                      & $P_0$ [days]                       & 11.56                            & 365.3                          & 11599 (31.78 yrs)          \\
Reference density               & $\rho_0$ [g cm$^{-3}$]            & $5.9 \times 10^{-4}$   & $5.9 \times 10^{-7}$   & $5.9 \times 10^{-10}$      \\
Reference surface density   & $\Sigma_0$ [g cm$^{-2}$]        & $8.9 \times 10^8$       & $8.9 \times 10^6$       & $8.9 \times 10^4$        \\
\hline  &                &                  &    Values in reference models at $\Sigma_d =3.0\times 10^{-2}$                 &                  \\
\hline
Initial density at $r=r_{\rm cav}$   & $\rho_{\rm d0}$     [g cm$^{-3}$]   & $2.4 \times 10^{-4}$  & $2.4 \times 10^{-7}$   & $2.3 \times 10^{-10}$  \\
Initial surface density                      & $\Sigma_{\rm d0}$   [g cm$^{-2}$] & $2.7 \times 10^7$      & $2.7 \times 10^5$        & $2.7 \times 10^3$ \\
\hline  &                &                  &    Projected values at $q_d =1.0\times 10^{-4}$           &                   \\
\hline
Initial density at $r=r_{\rm cav}$   & $\rho_{\rm dp}$     [g cm$^{-3}$]   & $8.0 \times 10^{-7}$  & $8.0 \times 10^{-10}$   & $8.0 \times 10^{-13}$   \\
Initial surface density                     & $\Sigma_{\rm dp}$   [g cm$^{-2}$] & $8.9 \times 10^4$      & $8.9 \times 10^2$   & $8.9$  \\
\hline \hline
\end{tabular}
\caption{\textit{Top rows:} reference values calculated for different
sizes of the disc-cavity boundary $r_{\rm cav}$. \textit{Middle rows:} 
 initial values of
density and surface density taken in the model. \textit{Bottom rows:} Projected values, where we took a small value of the surface density, $\Sigma_d=10^{-4}$.  
 \label{tab:units}}
\end{table*}

\subsection{Initial disc-cavity equilibrium }
\label{sec:initial}

We calculate the equilibrium distribution of density and pressure
in the disc and cavity using an approach described in \citet{RomanovaEtAl2019}. 
In this approach, it is suggested that the disc has a high density, $\rho_d$ while the cavity has very low
density, $\rho_{\rm cav}<<\rho_d$.  To support this configuration, we take a high temperature in the cavity and a low temperature in the disc,
such that at the disc-cavity boundary, the gas pressure in the disc equals the gas pressure in the cavity, $p_d=p_{\rm cav}$.

To construct our initial condition,  we first
determine the equilibrium in the equatorial plane. The initial
density distribution is given by
\begin{equation}
\rho(r, 0) = \left\{
\begin{array}{lr}
\rho_{\rm cav}  & {\rm if}\ r < r_{\rm cav} \\
\rho_{\rm d} \left(\frac{r}{r_{\rm cav}}\right)^{-n} & {\rm if}\ r
\ge \r_{\rm cav}
\end{array}
\right. \label{eqn_rhoinit}
\end{equation}
Here, $\rho_{\rm cav}$ is the density in the cavity and $\rho_{\rm d}$  is the disc density near the cavity boundary, $r=r_{\rm cav} $.
Parameter $n$ specifies
the radial profile of the disc density. We take a similar distribution for pressure in the disc and cavity.

We take a disc with semi-thickness $h/R=0.03$ (determined at $r=r_d$) and derive the temperature in the disc from the condition: $(h/R)_d=(c_s/v_K)_d$, where $c_s$ and $v_K$ are the sound speed and Keplerian velocities at $r=r_d$. Using our dimensionalization ($GM=1$, $r_0=r_{\rm cav}$, $v_{K0}=v_0=1$), we obtain the temperature at the inner edge of the disc: 
$T_d = c_s^2 =(h/R)^2=0.0009$. We determine the dimensionless density $\rho_d$ at  $r=r_d$ ($\rho_d=0.4$ in most of our simulations). We take much lower density in the cavity, $\rho_{\rm cav}=10^{-3} \rho_d$.

At the inner edge of the disc ($r=r_d$), we take an equal pressure for the disc and cavity in the equatorial plane: 
 $p_{\rm d}=p_{\rm cav}$. This condition provides a zero pressure gradient force at the boundary and the initial equilibrium between the disc and cavity.

The dimentionless
temperature is related to density and pressure by the ideal gas
law, ${\cal R} T(r) = p(r)/\rho(r)$. Therefore, $T_d \rho_d=T_{\rm cav} \rho_{\rm cav}$, and the temperature 
in the cavity $T_{\rm cav}=10^3 T_d$  is much higher than that in the disc\footnote{In protoplanetary discs, the temperature may be low in both the disc and the cavity (excluding cases where cavities are carved by stellar wind or high-energy radiation from the star).  We note that the condition of equal pressure (a high temperature in the cavity) is not a necessary conditions for the final equilibrium because the main forces that support the disc are gravitational and centrifugal forces, while the pressure gradient force is much smaller. }. 

Initially, the disc is isothermal. The temperature in the cavity is also constant. 
 Subsequently, at $t>0$, we
calculate the temperature distribution using the energy
equation.

In our model, the cavity has a very low density, $10^3$ times lower than in the disc.
 A planet in the low-density cavity excites the zero-order Lindblad resonances responsible for the migration. 
However,
the torques are proportional to the density (e.g., \citealt{GoldreichTremaine1978}) and are 
 $10^3$ times smaller than torques acting on a planet migrating inside the disc.
In simulations, we observed that a planet migrates inwards as long as some of Lindblad resonances are located inside the disc. 
However, migration stops when all resonances are located inside the cavity, and eccentric Lindblad resonances start acting.
The high temperature in the cavity does not influence the planet's  migration in the cavity \footnote{Thermodynamics, viscosity, and irradiation of the disc by a star
may change the direction of migration of a planet located in the disc (e.g., \citealt{BitschEtAl2013b,PierensRaymond2016}). However, 
in the low-density cavity, the influence of the high-temperature gas on the planet's migration is negligibly small.}.

Next, we assume that there is a hydrostatic equilibrium in the
vertical direction and build the 3D distribution of density:

\begin{equation}
\rho(r, z) = \rho(r,0)
\exp\left(\frac{\Phi(r,0)-\Phi(r,z)}{{\cal
R}T(r,0)}\right)~, \label{eqn_potential}
\end{equation}
 where $\Phi(r,z) = -GM_*/(r^2+z^2)^{1/2}$ is the
gravitational potential of the star. The expression for pressure
is analogous. The azimuthal velocity $v_{\phi}$ is determined from the
balance between gravity and pressure gradient forces in the radial
direction:
\begin{equation}
v_\phi(r,z) = \sqrt{r \left( \frac{\partial \Phi}{\partial
r} + \frac{1}{\rho} \frac{\partial p}{\partial r} \right)}~.
\label{eq:vphi}
\end{equation}
These formulae allow us to start from a quasi-equilibrium
configuration for the disc and the cavity.

We obtain the surface density distribution in the disc
 by integrating the volume density $\rho$ in the $z-$direction:
\begin{equation}
\Sigma=\int{\rho dz}\propto H \rho \propto \frac{c_s \rho}{\Omega}
\propto \frac{\sqrt{p \rho}}{\Omega} \propto r^{-s}~,  
\label{eqn:_Sigma}
\end{equation}
where, $c_s\propto\sqrt{p/\rho}$ is the sound speed
and $s={\frac{3-2n}{2}}$.

\subsection{Calculation of the planet's orbit.}

We calculate the orbit of the planet, taking into account the interaction of a planet with the star and the disc. 
We use the earlier developed approaches  (e.g. ,\citealt{Kley1998,Masset2000,KleyNelson2012,CominsEtAl2016,RomanovaEtAl2019}).
We find the position $\textbf{r}_{\rm p}$  (the radius vector from the star to
the planet) and velocity $\textbf{v}_{\rm p}$ of the planet at each time step solving
the equation of motion:
\begin{eqnarray}
 M_{p} \frac{d\textbf{v}_{\rm p}}{dt} = -\frac{GM_*M_{\rm p}}{|\textbf{r}_{\rm p}|^{3}}\textbf{r}_{\rm p}
                                                                              -\frac{GM_{\rm p}^{2}}{|\textbf{r}_p|^{3}}\textbf{r}_ {\rm p} + \textbf{F}_{\rm disc\rightarrow p} .
\label{eq:planet}
\end{eqnarray}
The first term on the right-hand side represents the gravitational force from
the star. The middle term accounts for
the fact that the coordinate system is centered on the star and is not inertial.
\begin{equation} 
\textbf{F}_{\rm disc\rightarrow p}=\int\frac{GM_{\rm p}}{|\textbf{r}-\textbf{r}_{\rm p}|^3}(\textbf{r}-\textbf{r}_{\rm p})\rho r dr d\phi dz
\end{equation} 
is a cumulative force acting from the disc to the planet.

We  calculate the planet's orbital energy and angular momentum
per unit mass using the calculated values of $\textbf{r}_{\rm p}$
and $\textbf{v}_{\rm p}$:
\begin{align}
    E_p = \frac{1}{2}\lvert{\textbf v}_{\rm p}\rvert^{2} - \frac{GM_*}{r_p}
    & & {\rm and}
    & &  {\textbf L_p} = \textbf{r}_{\rm p} \times \textbf{v}_{\rm p} .
\end{align}
We use these relationships to calculate the semi-major axis and eccentricity of the planet's orbit at each time step:
\begin{align}
    a_p = -\frac{1}{2}\frac{GM_*}{E_p} & & {\rm and}
    & &
    e_p = \sqrt{1 - \frac{L_p^{2}}{GM_* a_p}} .
\end{align}
We calculate the inclination angle of the orbit as
\begin{equation} 
i_p=\arccos{\bigg(\frac{L_{\rm zp}}{L_p}\bigg)} ~,
\end{equation} 
where $L_{\rm zp}-$is the $z-$component of the angular momentum.

\subsection{Evolution of the disc}

The evolution of the disc has been calculated using earlier-developed approaches 
\citep{KoldobaEtAl2016,RomanovaEtAl2019}.
Below, we briefly describe the numerical model.
We model the evolution of the accretion disc using 3D equations of
hydrodynamics:
\begin{equation}
\parti{\vec{U}}{t} + \nabla \cdot \vec{F}(\vec{U}) = \vec{Q} ~,
\label{eq:hydro}
\end{equation}
where 
$\vec{U}$ is the vector of conserved variables and
$\vec{F}(\vec{U})$ is the vector of fluxes:
\begin{equation}
\vec{U} = \left[\rho, \rho S, \rho \vec{v} \right]^T, \quad
\vec{F}(\vec{U}) = \left[\rho \vec{v}, \rho \vec{v} S, \tensor{M}
\right]^T, \label{eq:hydro_state}
\end{equation}
and $\vec{Q} = [0, 0, -\rho \nabla \Phi]$ is the vector of source
terms;
$\rho$ is the density, $\vec{v}$ is the
velocity vector, $S
\equiv p/\rho^\gamma$ is the entropy function,
 we take $\gamma = 5/3$ in all models, $\Phi$ is the gravitational potential of the
star-planet system  and
$\tensor{M}$ is the momentum flux tensor, with components $M_{ij} =
\rho v_i v_j + \delta_{ij} p  - \tau_{ij}~, $ where  $p$ is the fluid pressure,  and $\delta_{ij}$ is the
Kronecker symbol; $\tau_{ij}$ is the tensor of viscous stresses (we take into account only $r\phi$ and
$z\phi$ components). In our
code, we use the entropy balance equation instead of the full
energy equation.
We include a viscosity term, with the  viscosity coefficient in
the form of $\alpha-$viscosity, $\nu_{\rm vis}=\alpha c_s H$ \citep{ShakuraSunyaev1973}.  
 
The equations of hydrodynamics are integrated numerically using an
explicit conservative Godunov-type numerical scheme  \citep{KoldobaEtAl2016}.
At the external boundaries we use ``free" boundary conditions $\partial A/\partial r =0$ and
$\partial A/\partial z =0$ for all variables $A$. At these conditions, matter freely flows out of the simulation region. We place an additional condition: we forbid the inward flow of matter into the simulation region. At the
inner boundary, we use fixed boundary conditions. They provide better results compared with the free conditions. The density in the cavity is very low, 
and only an insignificant amount of matter accumulates at the boundary during the simulation run.
 We also use the procedure of damping waves
at the inner and outer boundaries, following
\citet{FromangEtAl2005}. In addition, we place an exponential cut to the density distribution at the radius $r=0.7 R_o$ (with the width of exponential decay of 
$0.2 R_o$)
to be sure 
that the external boundary does not influence the result.

The simulation region represents a flat cylinder that stretches in the radial direction between the inner and outer boundaries,   $0.3<R<18$,
and in the vertical direction  between values of  $-1.5<z<1.5$.
The grid is evenly
spaced in the azimuthal and vertical directions, where the number
of grid cells in most of the simulations is $N_\phi=300$ and $N_z=72$, respectively. 
In the radial direction, $N_r=168$, and the size of grids increases with the distance from the star such that the grids have
approximately a square shape.  The code is parallelized using MPI. Typical number of processors per one simulation run is 280.

\begin{figure*}[h]                                                                                                                                                                                                                                                    
     \centering
     \includegraphics[width=0.8\textwidth]{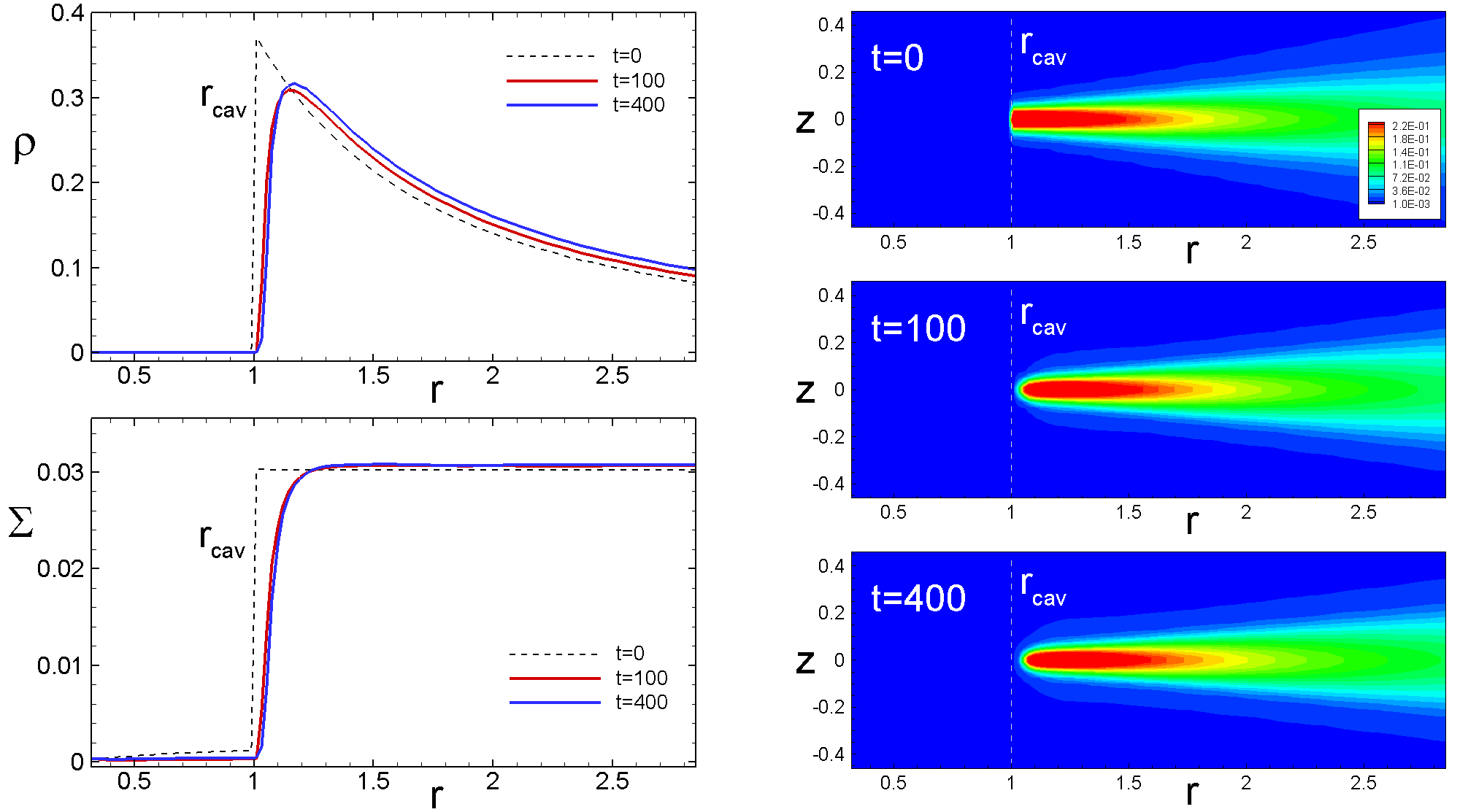} 
     \caption{\textit{Left panel:} Linear distribution of the density $\rho$ and surface density $\Sigma$  in the inner disc  in models with zero planet mass $m_p=0$ at  $t=0$  (dashed line) and  $t=100, 400$.  \textit{Right panel:} $rz$ slices of the density distribution $\rho$, at the same moments in time.     
\label{fig:1d-xz-m0}}
\end{figure*}

\begin{figure*}[h]                                                                                                                                                                                                                                                    
     \centering
     \includegraphics[width=0.8\textwidth]{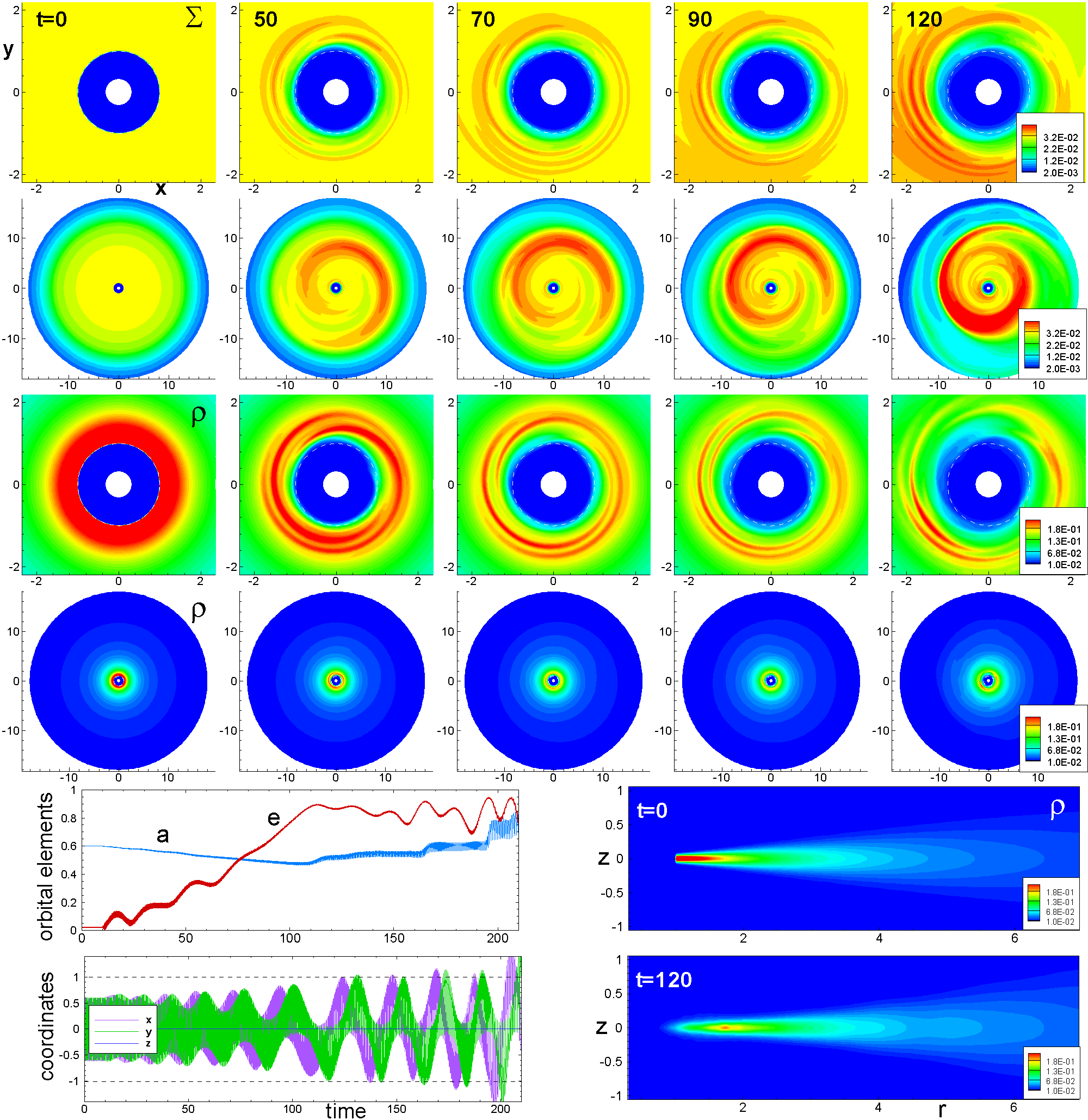} 
     \caption{Simulation results of the disc-planet interaction in model $n1.5i0$. The top row of panels shows the surface density distribution $\Sigma$  in the inner part of the simulation region at different moments in time, $t$. The 2nd row shows the same but in the whole simulation region.  The 3rd and 4th rows show the same but for the volume density distribution $\rho$ in the equatorial plane.  
The blue and red colours show the lowest/highest values of surface density and density.  
\textit{Bottom left panels}: 
Top: temporal evolution of the semi-major axis, $a$, and eccentricity, $e$. Bottom: temporal evolution of the planet's coordinates, $x, y, z$. The time is measured in periods of Keplerian rotation at the initial location of the cavity boundary,  $r_{\rm cav}=1$. \textit{Bottom right panels:} density distribution in the $rz-$plane at $t=0$ and $t=120$.  
\label{fig:n1.5-all}}
\end{figure*}

\begin{figure*}[h]                                                                                                                                                                                                                                                    
     \centering
     \includegraphics[width=0.8\textwidth]{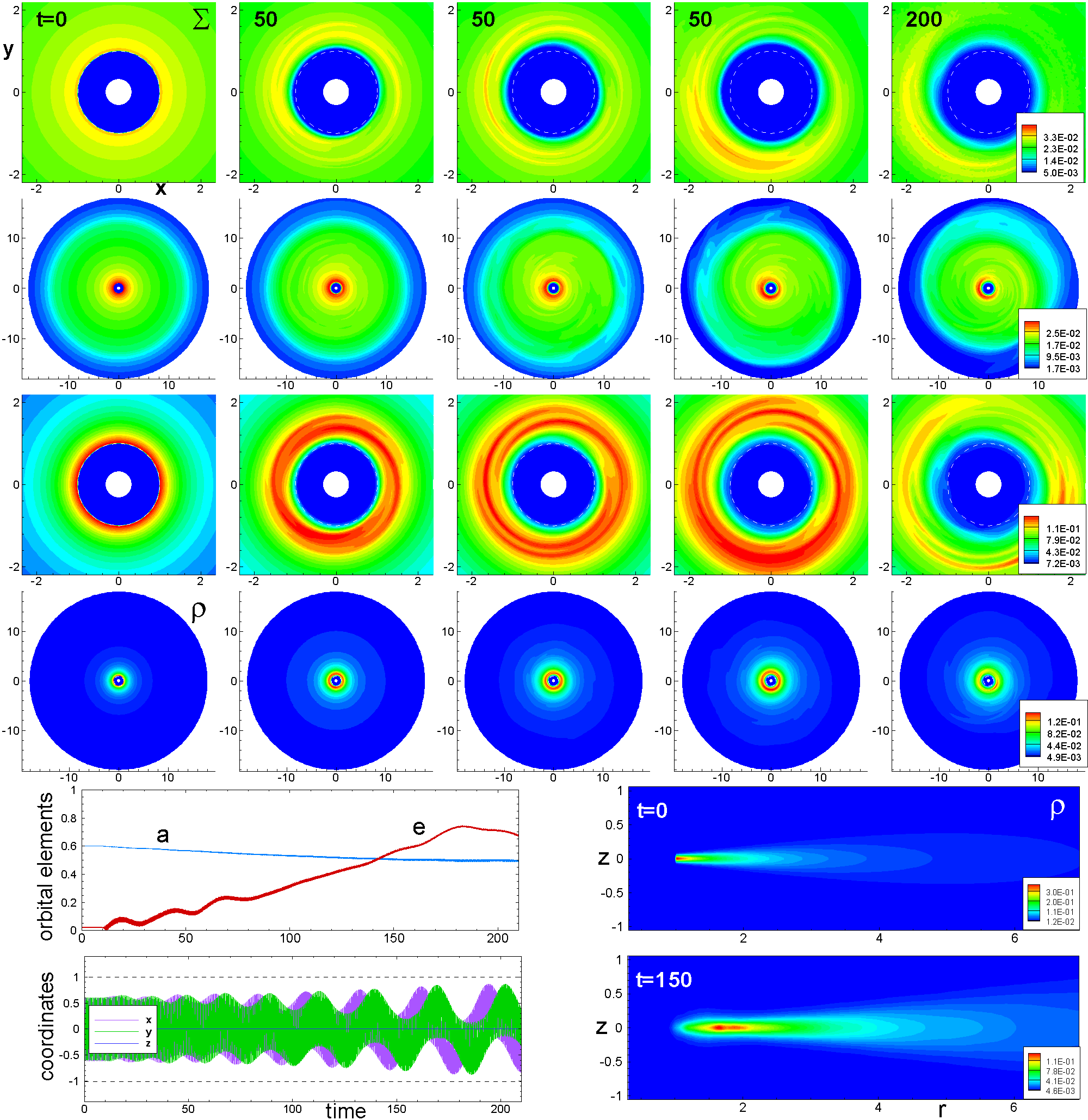} 
     \caption{Same as in Fig. \ref{fig:n1.5-all} but  for reference model $n1.8i0$. Note that we use slightly different color bars at different scales to show the fine structure of ELRs
in enlarged plots and features in the whole disc in expanded plots.
\label{fig:n1.8-all}}
\end{figure*}

\subsection{Reference models}

In the reference models, we take a planet of mass $M_p=5 M_J$ ($q_p=5\times 10^{-3}$) and a disc with 
a reference surface density $\Sigma_d=3\times 10^{-2}$,   viscosity coefficient $\alpha=3\times 10^{-4}$ and semi-thickness of the disc $h=H/r=0.03$ (determined at the inner edge of the disc, $r=r_{\rm cav}$). 
We take two values of slopes in the equatorial density distribution $n=1.5$ and $1.8$
which correspond to slopes in the surface density distribution: $s=0$ and $s=0.3$, respectively.
We place a planet in the orbit with a semi-major axis $a_0=0.6$ (see Tab.
 \ref{tab:ref-model} for parameters in Reference model).
 In test models of R23, we observed that in models with $0.6<a_0<1.0$, the result is very similar. However, initially,
a planet migrates inward due to the principal 1:2 outer Lindblad resonance (OLR), and only later, when $a_0\approx 0.6$, the eccentricity starts to increase due to 1:3 ELR. That is why,
to save computing time, we place a planet at $a_0=0.6$.  

We take non-zero initial eccentricity, $e_0=0.02$. In test models of R23 with zero eccentricity,
there is an interval of time where the eccentricity increases, but slowly, due to the opposing action of the  1:2 eccentric corotation resonance (ECR), which damps the eccentricity. However,
the ECR resonance is saturated in models with a small initial eccentricity (e.g., \citealt{GoldreichSari2003,OgilvieLubow2003}). To save computing time,   we take $e_0=0.02$ in all our models. 

We take the inclination angle of the orbit $i_0=0$ in models where a planet is located in the plane of the disc and different values of $i_0$ in models where we investigate inclined orbits. We name reference models using parameters $n$ and $i_0$ ($n1.5i0$, $n1.8i0$, $n1.5i45$, etc.) In these models, we take common parameters shown in Tab. \ref{tab:ref-model}. If we change other parameters, then we specify them in the text and supplementary tables.

\begin{table}
\begin{tabular}[]{ c  | c }
 \hline
                        Parameter                                 &       Value                                          \\
\hline          
                mass of the planet (in $M_\odot$)            &    $q_p=5\times 10^{-3}$       \\
             mass of the planet in Jupiter mass               &    $M_p=5$                             \\
                 semi-thickness of the disc                          & $h=0.03$                              \\
                reference surface density                             &    ${\tilde{\Sigma}}=3\times 10^{-2}$    \\
               coefficient of viscosity                                  & $\alpha=3\times 10^{-4}$                   \\
               slope in $\rho$ distribution                          &   $n=1.5$ ~ or ~  n=1.8    \\
                semi-major axis of the planet               &    $a_0=0.6$                                     \\ 
                initial eccentricity                                        &   $e_0=0.02$                              \\
     inclination of planet's orbit                 &  ~ $i_0=0^\circ$ ~or~ $i_0=45^\circ$  \\
\hline 
\end{tabular}
\caption{Parameters in main reference models. We name reference models using parameters $n$ and $i_0$: $n1.5i0$, $n1.8i0$,
$n1.5i45$, $n1.8i45$, etc.
  \label{tab:ref-model}}
\end{table}

\subsection{Testing initial equilibrium}

First, we take a planet with zero mass $m_p=0$ and test the stability of our initial conditions determined by equations (1)-(4).
Fig. \ref{fig:1d-xz-m0} compares density distributions at different moments in time. Right panels show $rz$ projections of the density distribution
at $t=0$ (top panel) and at moments $t=100$ and $t=400$. The left panels show the radial distribution of the equatorial density $\rho$ (top) 
and surface density $\Sigma$. One can see that initially, the density distribution changed slightly, but at later times, it stayed approximately the same. 
Therefore, 
the initial conditions described in Sec. \ref{sec:initial} provide a good equilibrium between matter in the disc and cavity.  The simulation time of 
$t=400$ is longer than the duration of our simulation runs. We conclude that this initial setup can be used as a base for our 3D simulations.
These initial conditions may directly describe inner cavities around the star formed by the magnetosphere, stellar wind, or high-energy radiation from the star.  Cavities at larger distances from the star may have dust, pebbles, and regions of forming planets.
In cavities, located away from the star, temperature can be low. In our model, the density in the cavity is so low that neither density nor high temperature influence the dynamics of the planet. Planet interacts gravitationally only with a star and the disc.

\section{Planets at aligned orbits}
\label{sec:aligned}

First, we take a reference model $n1.5i0$  with typical reference parameters (see Tab. \ref{tab:ref-model})  and  place an orbit of the planet in the equatorial plane of the disc (at inclination angle $i_0=0$). We keep a planet in a fixed orbit during the first 10 rotations of the inner disc and then release it. 
The 1st row from the top of Fig. \ref{fig:n1.5-all} shows the  surface density distribution, $\Sigma$, in the inner part of the simulation region, 
near the cavity.
 One can see two tight spiral waves  at $t=50$ and three waves at $t=70, 90, 120$.  These waves are similar to waves observed in 2D simulations of R23. They correspond to 
$m=2$ and $m=3$ modes of the ELR. They drive the eccentricity of the planet up to $e\approx 0.8$. 

The 2nd row from the top shows the surface density distribution in the whole simulation region.   One can see that a one-armed density wave forms in the inner disc and propagates to large distances.  

The 3rd raw shows that   the $m=2$ and $m=3$ resonant ELR density waves also present in the equatorial density distribution. The 4th row shows the equatorial density distribution in the whole region.

The bottom left panels of the same figure show the temporal evolution of orbital parameters (top) and coordinates of the planet (bottom).
 One can see that the eccentricity increases during $t\approx 120$, reaches $e\approx 0.8$, and then varies in the interval of $0.7\lesssim e\lesssim 0.9$. The semi-major axis initially decreases and then increases. 

 We use the Cartesian coordinates to track the position of the planet, where coordinates $x$ and $y$ are located in the equatorial plane of the disc, and the $z-$coordinate coincides with the $z-$coordinate of the cylindrical system. One can see that $x$ and $y$ coordinates increase, and at $t\approx 120$, they reach the inner parts of the disc-cavity boundary at $r\approx 1-1.1$.
Oscillations of x and y coordinates reflect precession of the planet's orbit with a typical time scale of $t_{\rm prec}\approx 25-30$ (see more details in Sec. \ref{sec:disc-ecc}).

The bottom right panels show the $rz-$ slices of the density distribution at $t=0$ and $120$. One can see that during the simulation, the disc  changed its structure due to interaction with the planet. Changes are more significant than in the zero-mass planet's model (see Fig. \ref{fig:1d-xz-m0}). However, the low-density cavity is present, and the disc can be used to investigate the current problem. 

 When the planet reaches the inner parts of the disc, the eccentricity starts decreasing due to the coorbital corotation torque. At the cavity boundary, this torque is asymmetric and  tends to move a planet to larger radii (e.g., \citealt{MassetEtAl2006, RomanovaEtAl2019}). This may explain why the semi-major axis increases after $t\approx 120$. 
The ELR resonances tend to move to larger radii. However, a planet also moves to larger radii. This provides quasi-stationary situation when the eccentricity is high for a while. On the other hand, ELR resonances in the disc become non-axisymmetric, and precession of matter in the inner disc may gradually smear resonances. This also stops the planet's eccentricity growth  \footnote{A planet continues interacting with the disc gravitationally and may exchange its eccentricity with the disc.   
\citet{RagusaEtAl2018} performed long-lasting 2D simulations of the disc-planet interaction at late states of evolution and  low masses of the disc, and observed that a planet and a disc exchange eccentricity quasiperiodically (and in antiphase) for a long time.}. 
The final eccentricity of the planet depends on the particular situation, such as the rate of the disc dispersal.

Fig. \ref{fig:n1.8-all} shows the same values but in reference model $n1.8i0$ where the surface density decreases with the distance as $\Sigma\sim r^{-0.3}$. One can see that similar ELR resonances are observed in the inner disc. However, the one-armed density wave is weaker. 

\subsection{Dependence on parameters}
\label{sec:dependencies}

We take reference models $n1.5i0$ and $n1.8i0$ and investigate the eccentricity growth at different parameters. 

\subsubsection{Dependence on the mass of the planet} 

\begin{figure*}[h]
     \centering
     \includegraphics[height=0.28\textwidth]{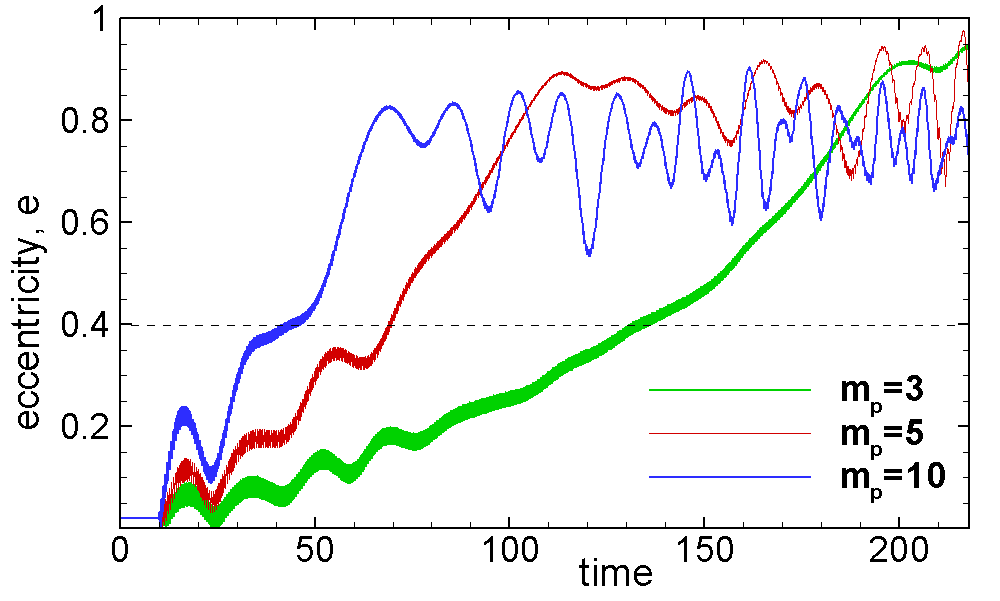} 
     \includegraphics[height=0.28\textwidth]{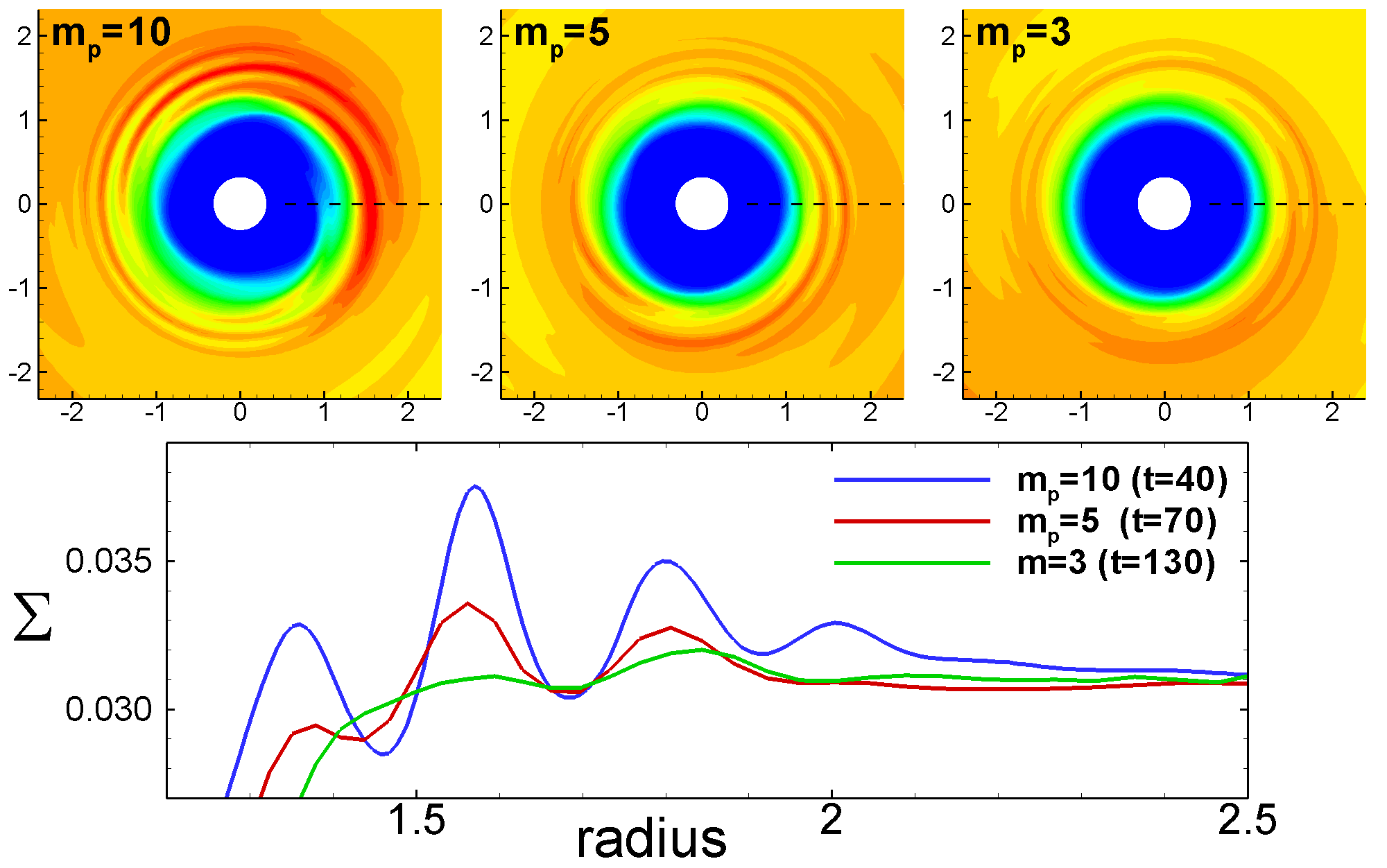} 
     \caption{\textit{Left panel:} Eccentricity evolution in the model $n1.5i0$ at different masses of the planet, $m_p$.  The time is measured in periods of Keplerian rotation at the initial location of the cavity boundary,  $r_{\rm cav}=1$. \textit{Right panels:} Top: the surface density distribution for these models at moments when $e\approx 0.4$. Bottom: The surface density distributions in the equatorial plane of the above figures along $x-$direction  (marked in dashed lines).
\label{fig:n1.5-mass}}
\end{figure*}

We performed simulations 
at the lower, $m_p=3 M_J$ and higher, $m_p=10 M_J$ mass of the planet. Left panel of Fig. \ref{fig:n1.5-mass} shows that in both models, the eccentricity increased up to $e\approx 0.8-0.9$. However, the eccentricity growth rate is smaller/larger in models with a smaller/larger planet mass. The right panels of the same figure show that the amplitude of ELR density waves is larger in models with larger planet mass.
These simulations show that the torque increases with the planet's mass, as predicted in theory.

\begin{table*}
\begin{tabular}[h]{ c  }
\hline
 Testing models with aligned orbits ($i_0=0$) at different parameters\\
\end{tabular}
\begin{tabular}[h]{ c  | c | c|c|c|c}
\hline          
\noindent Mass of the planet & $m_p (M_J)$         &    $0$   &   $3$                          &  $5$                      &  $10$            \\   
\hline
\noindent Viscosity       &  $\alpha-$parameter       &  $3\times10^{-4}$      &  $1\times 10^{-3}$ & $3\times 10^{-3}$ &  $1\times 10^{-2}$     \\  
\hline
 Grid resolution           &  $N_\phi$                       &   $100$             &   $200$      &  $300$       &    $400$ \\  
\hline
Slope in the density distribution &  $n$             &      $1.5$   &  $1.8$      &  $2$       & $2.5$ \\   
\hline
\end{tabular}
\caption{A set of parameters used to test a reference model $n1.5i0$.    
  \label{tab:mass}}
\end{table*}

\subsubsection{Dependence on viscosity}

\begin{figure*}
     \centering
     \includegraphics[width=0.39\textwidth]{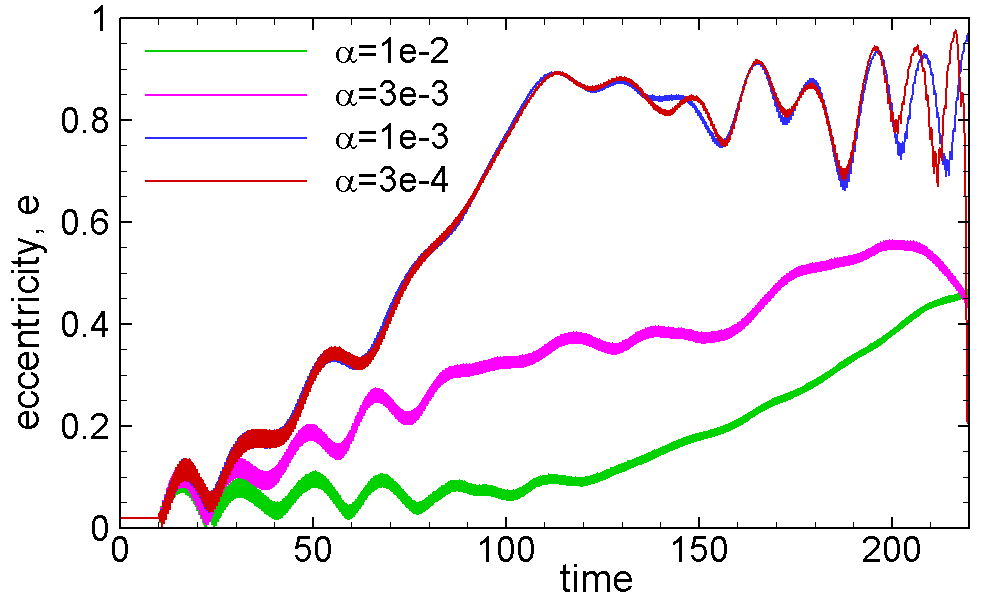} 
   \includegraphics[width=0.6\textwidth]{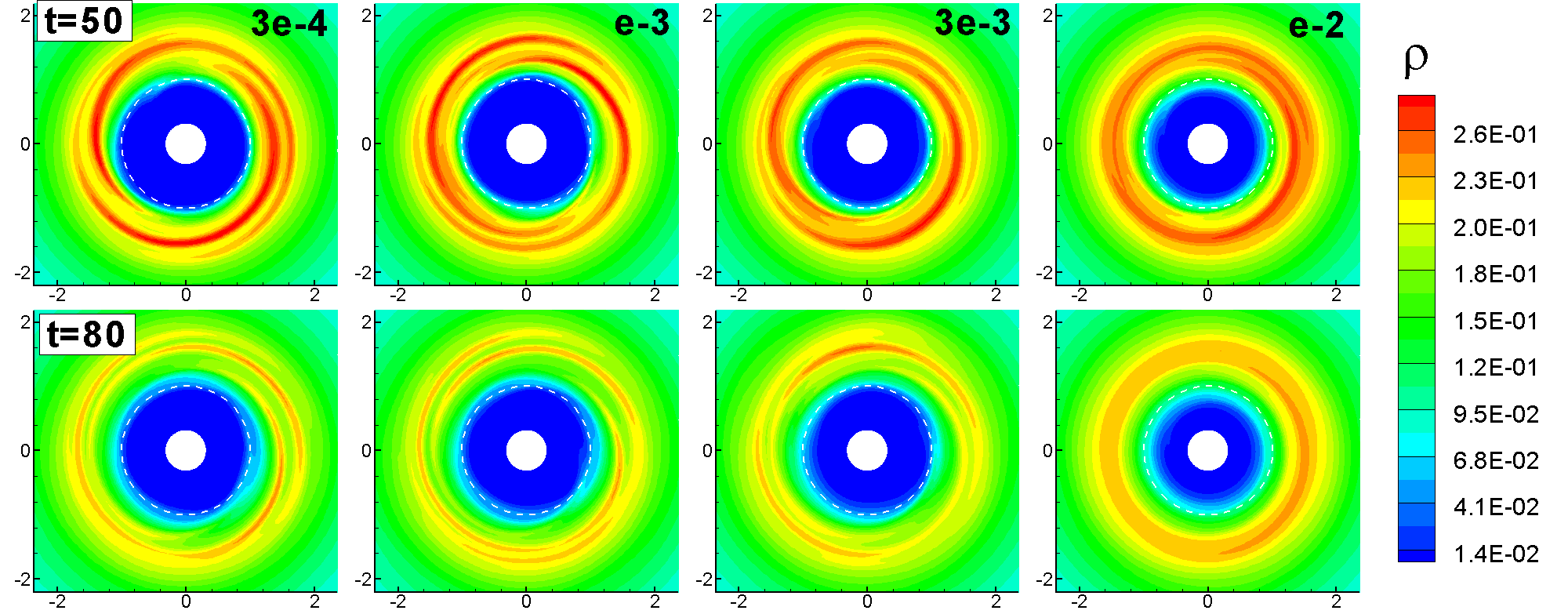} 
     \caption{\textit{Left panel:} Eccentricity evolution in model $n1.5i0$ but with different values of the $\alpha-$parameter of viscosity in the disc. \textit{Right panels:} Equatorial density distribution at different values of $\alpha-$parameter at times $t=50$ (top panels) and $t=80$ (bottom panels).
\label{fig:n1.5-visc-grid}}
\end{figure*}

 We performed test simulations at several values of the $\alpha-$parameter of 
viscosity from $\alpha=10^{-2}$ up to  $\alpha=3\times 10^{-4}$. 
The left panel of Fig.  \ref{fig:n1.5-visc-grid} shows the eccentricity variation at different $\alpha$.
The results are almost the same at $\alpha=10^{-3}$ and $\alpha=3\times 10^{-4}$. However,
the eccentricity increases slower when $\alpha=3\times 10^{-3}$ and even slower when $\alpha= 10^{-2}$.
Right panels show that the ELR density waves becomes more smeared in models with higher values of $\alpha$ taken at moments $t=50$
(top panels) and $t=80$ (bottom panels).
This result is similar to our 2D simulations (R23). Both types of simulations show that  the action of
ELRs decreases at higher viscosities in the disc.

\begin{figure*}
     \centering
   \includegraphics[width=0.39\textwidth]{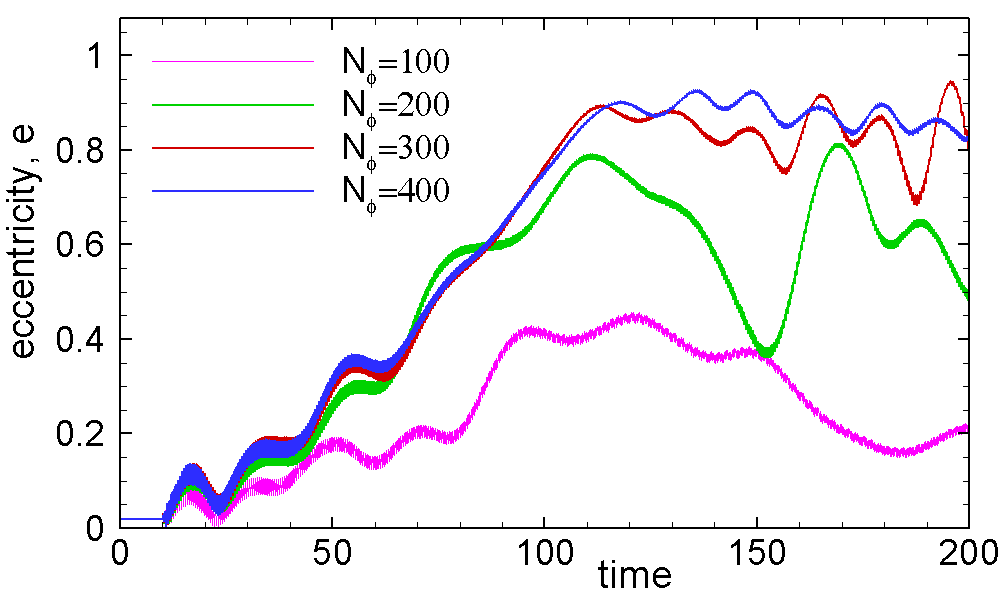} 
  \includegraphics[width=0.6\textwidth]{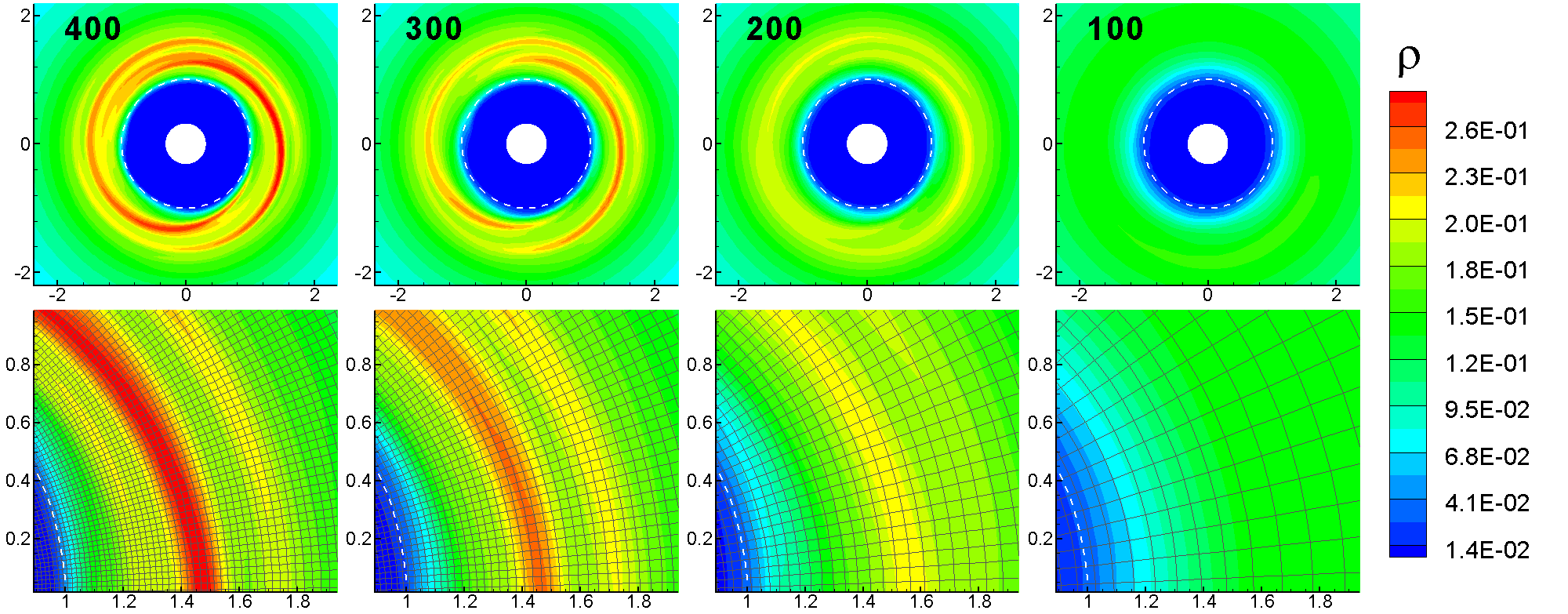} 
     \caption{\textit{Left panel:}  Eccentricity evolution in models with parameters of the reference model $n1.5i0$ but with different
grid resolutions $N_\phi=400, 300, 200, 100$.  \textit{Right panels:} \textit{Top:} Equatorial density distribution at $t=50$ and at different grids. 
\textit{Bottom:} the same, but a part of the region and the grid are shown.   
\label{fig:n1.5-grid}}
\end{figure*}

\begin{figure*}
     \centering
     \includegraphics[width=0.8\textwidth]{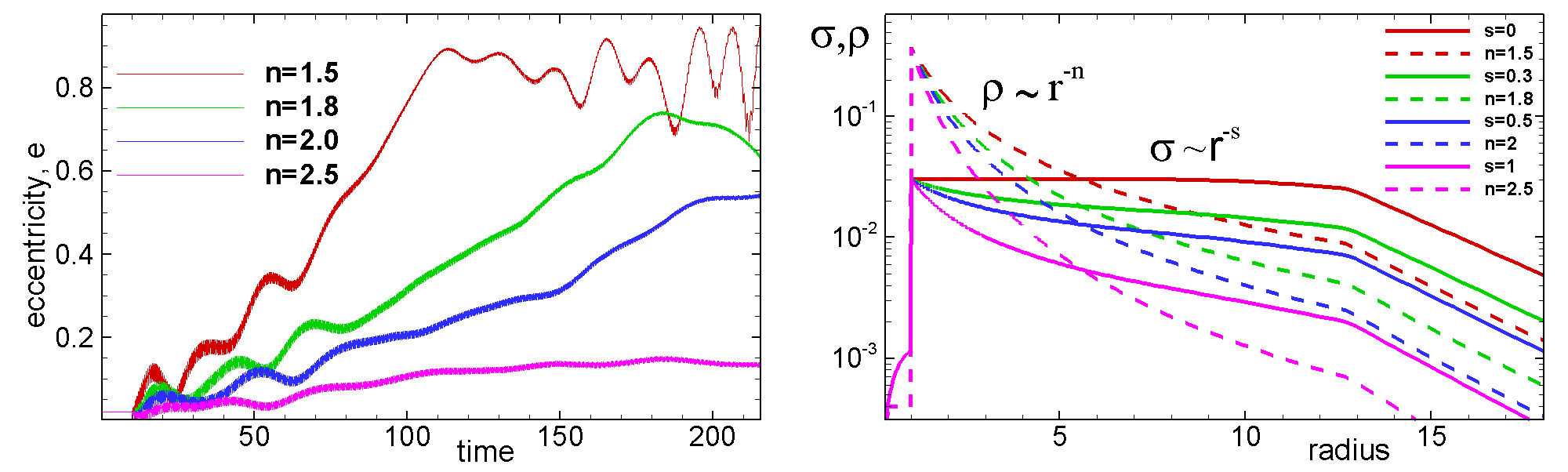} 
     \caption{\textit{Left panel:} Eccentricity evolution in models with different slopes $n$  in the equatorial density distributions $\rho\sim r^{-n}$. \textit{Right panel:} 
Initial equatorial density distribution and surface density distribution, $\Sigma\sim r^{-s}$, with radius in models with different $n$.  
\label{fig:ecc-an-2}}
\end{figure*}

\subsubsection{Dependence on the grid resolution}

The grid resolution taken in our models is $N_r\times N_\phi \times N_z = 168\times 300\times 72$.
For comparison, we calculated models with a lower and higher grid resolutions: $70\times  100\times 24$,  $119\times  200\times 48$, and $217\times  400\times 96$.

 Left panel of Fig.  \ref{fig:n1.5-grid} shows  that   at the higher grid resolution  ($N_\phi=400$), the curve for eccentricity evolution almost coincides with that for 
$N_\phi=300$ grid. However, the simulations are 2.6 times slower.  At the lower grid resolution ($N_\phi=100$ and $200$)  the simulations are faster. However, the eccentricity increases increases slower or decreases  (see pink and green lines in the figure). 
That is why we chose the grid $N_\phi=300$ in our simulations.

Both 2D and 3D simulations show  the necessity of high grid resolution while modeling ELRs. Models with a low 
 grid resolution may show a lower eccentricities compared with the higher grid resolution and  hence may 
underestimate the final value of the planet eccentricity.

Right panels of Fig. \ref{fig:n1.5-grid} show the equatorial density distribution after $t=50$ rotations in simulations with different grids. The bottom panels show
a part of the simulation region and the grid. One can see that at grid resolutions $N_\phi=400$ and $300$, the ELR density waves are well resolved (with many grids across the wave)
and are of high density (amplitude, red color).  At the lower grid, $N_\phi=200$, the density waves are resolved only by a few grid cells, and the amplitude of waves is smaller (yellow color). At even a lower grid resolution,
$N_\phi=100$, the grid does not resolve the wave, and we do not observe the wave.  We think that at lower grid resolutions, the numerical diffusivity is high and the denser matter of spiral waves diffuses away from their initial positions.


We should note that at the lower density of the inner disc and lower mass
of the planet, the torques are weaker, and the grid resolution should be higher to resolve ELRs. For example, in R23, at the low disc density, $q_d=\Sigma_d=3\times 10^{-4}$, the grid resolution of $N_\phi=600$ was necessary to resolve ELRs. However,
in current 3D simulations where $\Sigma_d=3\times10^{-2}$, the disc density is high,  and the grid with $N_\phi=300$ is sufficient.

\begin{figure*}[h]
     \centering
     \includegraphics[width=0.8\textwidth]{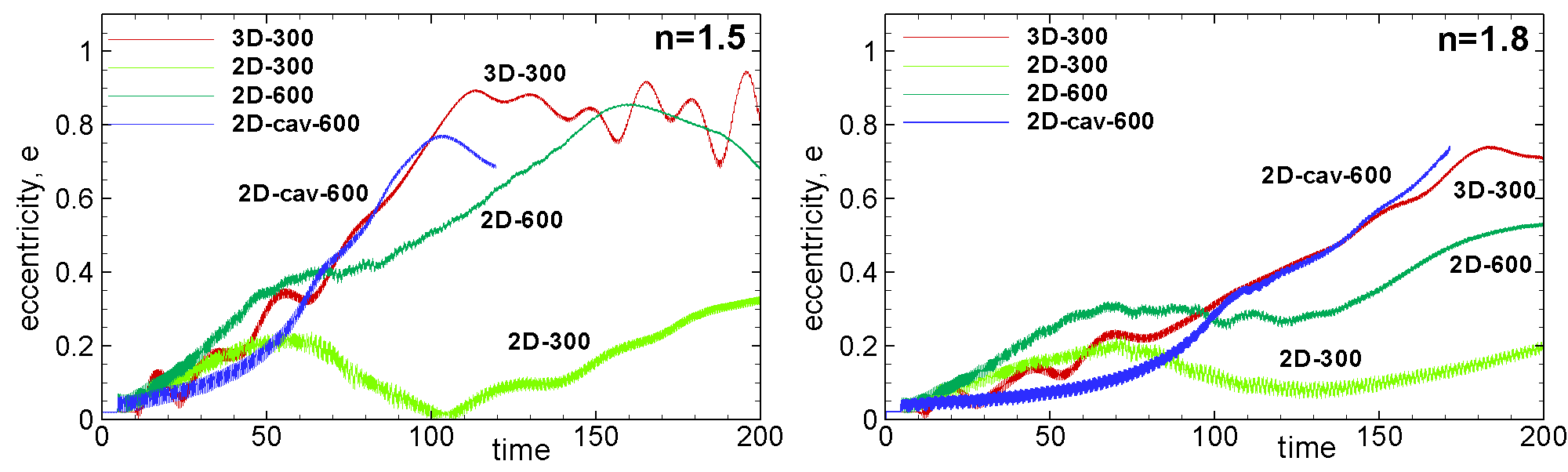} 
     \caption{\textit{Left panel:} Eccentricity evolution in 3D model $n1.5i0$ (3D-300)and 2D models with similar parameters. 2D models were calculated at grids $N_\phi=300$ (2D-300) and $N_\phi=600$ (2D-600). A model 2D-cav-600 corresponds to the fixed 
cavity model of R23 taken at the grid resolution of $N_\phi=640$ (see also Tab. \ref{tab:2D-3D}).
 \textit{Right panel:} The same but for the model $n1.8i0$.   
\label{fig:3d-2d-2}}
\end{figure*}

\subsubsection{Dependence on the slope of the density distribution }

 We compared the eccentricity evolution at different initial slopes of the density distribution in the disc. We took 
discs with equatorial density distributions  $\rho\sim r^{-n}$
with $n=1.5, 1.8, 2, 2.5$. They correspond to the surface density distributions:
 $\Sigma\sim r^{-s}$, with $s=0, 0.3, 0.5, 1$. The right panel of Fig. \ref{fig:ecc-an-2} shows the initial density and surface density distributions with radius. The left panel shows that the eccentricity increases slower in models with steeper slopes in density distribution. We think that the eccentricity growth rate decreases with $n$ because at steeper density distributions, the inner disc mass $\Sigma_{\rm d0} r_0^2$ and the total mass of the disc is smaller.
When we increase the density 1.5-2 times, we observe faster eccentricity growth.  After multiple experiments with different $s$, we conclude that the steeper density distribution 
is not a factor that may stop eccentricity growth.

\subsubsection{Comparison of 3D and 2D simulations}

We compared the eccentricity evolution in our reference 3D models $n1.5i0$ and $n1.8i0$ with 2D models calculated at the same  physical parameters (see Tab. \ref{tab:ref-model}) and the grid.  
We also compared 3D models with a 2D model of R23 where the cavity radius has been fixed and simulations of the gas flow were performed only at the radius $r > r_{\rm cav}$. 
The left and right panels of Fig. \ref{fig:3d-2d-2} compare the eccentricity evolution in 3D and 2D simulations in models with $n=1.5$ and $n=1.8$. One can see that in 2D model 2D-300, the eccentricity increases with the same rate as in the 3D model up to $e\approx 0.2$. Subsequently,  it decreases. The eccentricity increases again but does not reach high values, like in the 3D model. In test 2D simulation with a higher grid resolution, $N_\phi=600$ (model 2D-600), the eccentricity increases faster and reaches higher values than in model 2D-300. 
In the model with the fixed cavity
(2D-cav-600) calculated using R23 approach but at the same parameters as in 3D models, the eccentricity initially  increases slower than in 3D, but subsequently, it grows to high values, like in the 3D model (see blue lines in Fig.  \ref{fig:3d-2d-2}).  These simulations stopped when the planet reached the cavity boundary.
So, we observed that 2D models with non-fixed boundaries show slower eccentricity growth than 3D models.
This phenomenon may be due to the faster eccentricity growth in 3D models, discussed by \citet{TeyssandierOgilvie2016}. They found that the eccentricity growth rate due to ELRs is 2-4 orders of magnitude larger in 3D models than in 2D models (see growth rate in adiabatic models in their Table 6). This issue should be studied separately.  

\begin{table*}
\begin{tabular}[]{ c  | c | c | c | c }
 \hline
 Model            &   3D-300                         & 2D-600     &  2D-300  &  2D-cav-600       \\
\hline          
grid/dimension                   &   3D                                 & 2D           &  2D          &  2D (fixed cavity)  \\
\hline   
  $N_\phi$                       &   300                               &  600         &  300       &  640                  \\   
  $N_r$                           &  168                                 & 406         & 203      & 336              \\
   $N_z$                          &   72                                  & --            &      --      & --               \\
\hline
\end{tabular}
\caption{Models used for comparisons of the 3D and 2D models. See comparisons in Fig. \ref{fig:3d-2d-2}.
  \label{tab:2D-3D}}
\end{table*}

\subsubsection{Time scales of eccentricity growth} 

From the left panel of  Fig.  \ref{fig:3d-2d-2}, we estimate the eccentricity growth rate in 3D simulations:
 $t_{\rm ecc}^{-1}=d(\ln{e})/dt\approx 0.024$, and the time scale  $t_{\rm ecc}\approx 43.48$. 
This value
is relevant to our reference model, where $\Sigma_d=3\times 10^{-2}$.  Our earlier 2D simulations
performed in the range of  $10^{-2}<\Sigma_d<10^{-4}$  have shown that $t_{\rm ecc}^{-1}\sim \Sigma_d$ and also $t_{\rm ecc}^{-1}\sim m_p$, which is in accord with theoretical studies (e.g., \citealt{GoldreichTremaine1978}).
Here, we  project the time scale of the eccentricity growth to more realistic parameters of $\Sigma_d=10^{-4}$. We also convert time to dimensional units, taking into account that we measure time in units of $P_0=2\pi r_0/v_0=2\pi r_{\rm cav}^{3/2}/\sqrt{GM_\odot}$. We take $r_{\rm cav}=10$ AU as a reference scale (see Tab. \ref{tab:units}). 
We obtain the dimensional time of eccentricity growth as:
\begin{equation}
T_{\rm ecc}\approx 4.14\times 10^5 yr \bigg(\frac{r_{\rm cav}}{10AU}\bigg)^{3/2}\bigg(\frac{\Sigma_d}{10^{-4}}\bigg)^{-1}\bigg(\frac{m_p}{5M_J}\bigg)^{-1}~.
\label{eq:t-ecc}
\end{equation}  
Eccentricity will grow if the cavity does not change its position significantly or the disc does not disperse. For example, if the cavity is present  during $T_{\rm cav}=10^6$ years, then  the eccentricity will increase significantly, if $T_{\rm ecc}\lesssim T_{\rm cav}$, or if the cavity radius 
\begin{equation}
r_{\rm cav} \lesssim 18 {\rm AU} \bigg[\bigg(\frac{T_{\rm cav}}{10^6{\rm yr}} \bigg)\bigg(\frac{\Sigma_d}{10^{-4}}\bigg)\bigg(\frac{m_p}{5M_J}\bigg)\bigg]^{2/3}~.
\end{equation} 
Planets in a small-sized cavity have a high rate of eccentricity growth. They may have several episodes of eccentricity growth (due to the above-discussed mechanisms) and decay due to local corotation torque when the planet starts entering the inner disc.
\begin{figure*}
     \centering
     \includegraphics[width=0.8\textwidth]{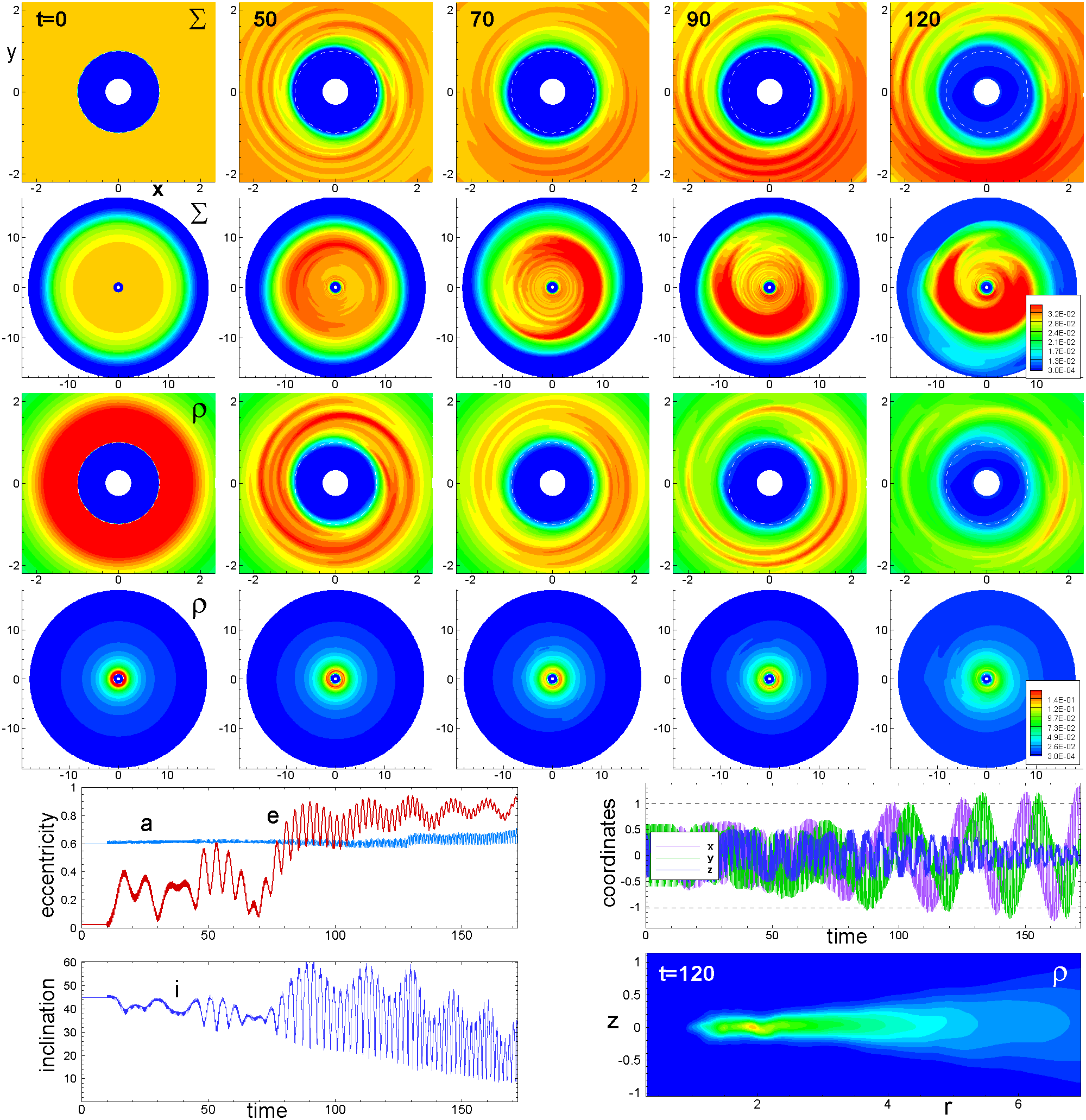} 
     \caption{
{\textit Top 4 rows:} The same as in Fig. \ref{fig:n1.5-all} but for the model $n1.5i45$. 
\textit{Bottom left panels}: 
Top: temporal evolution of the semi-major axis, $a$, and eccentricity, $e$. Bottom: temporal evolution of the inclination of the orbit, $i$. 
 \textit{Bottom right panels:} Top: temporal evolution of the planet's coordinates, $x, y, z$. Bottom: density distribution in the $rz-$plane at $t=120$.  
\label{fig:n1.5-45-all}}
\end{figure*}

\begin{figure*}
     \centering
     \includegraphics[width=0.8\textwidth]{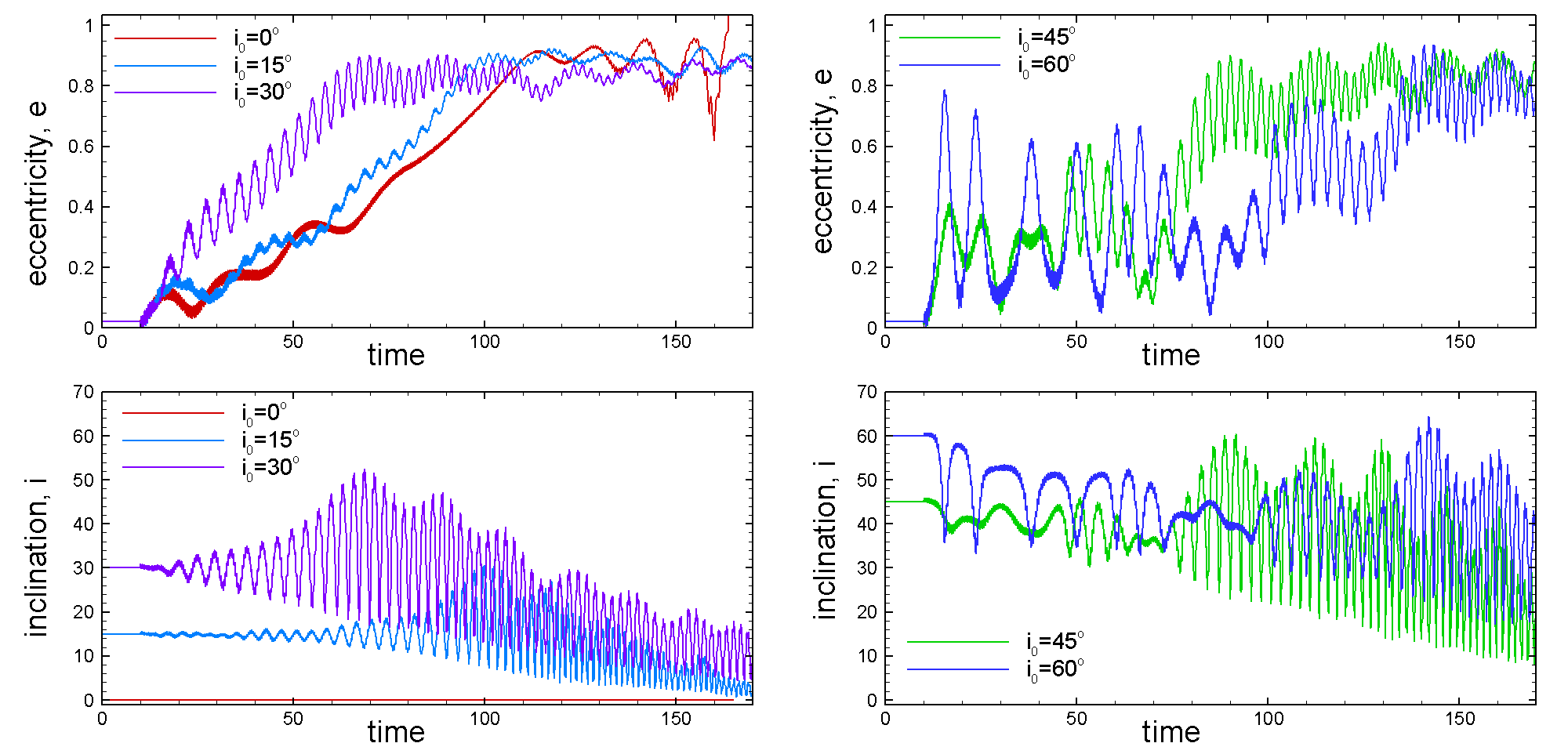} 
     \caption{{\it Left panels:} Temporal evolution of the eccentricity, $e$ and  inclination, $i$ in models with $n=1.5$ and initial values of the orbit inclination: $i_0=0^\circ,15^\circ,30^\circ$.
{\it Right panels:} The same but for $i_0=45^\circ, 60^\circ$.
\label{fig:n1.5-ei-4}}
\end{figure*}

\begin{figure*}
     \centering
     \includegraphics[width=0.8\textwidth]{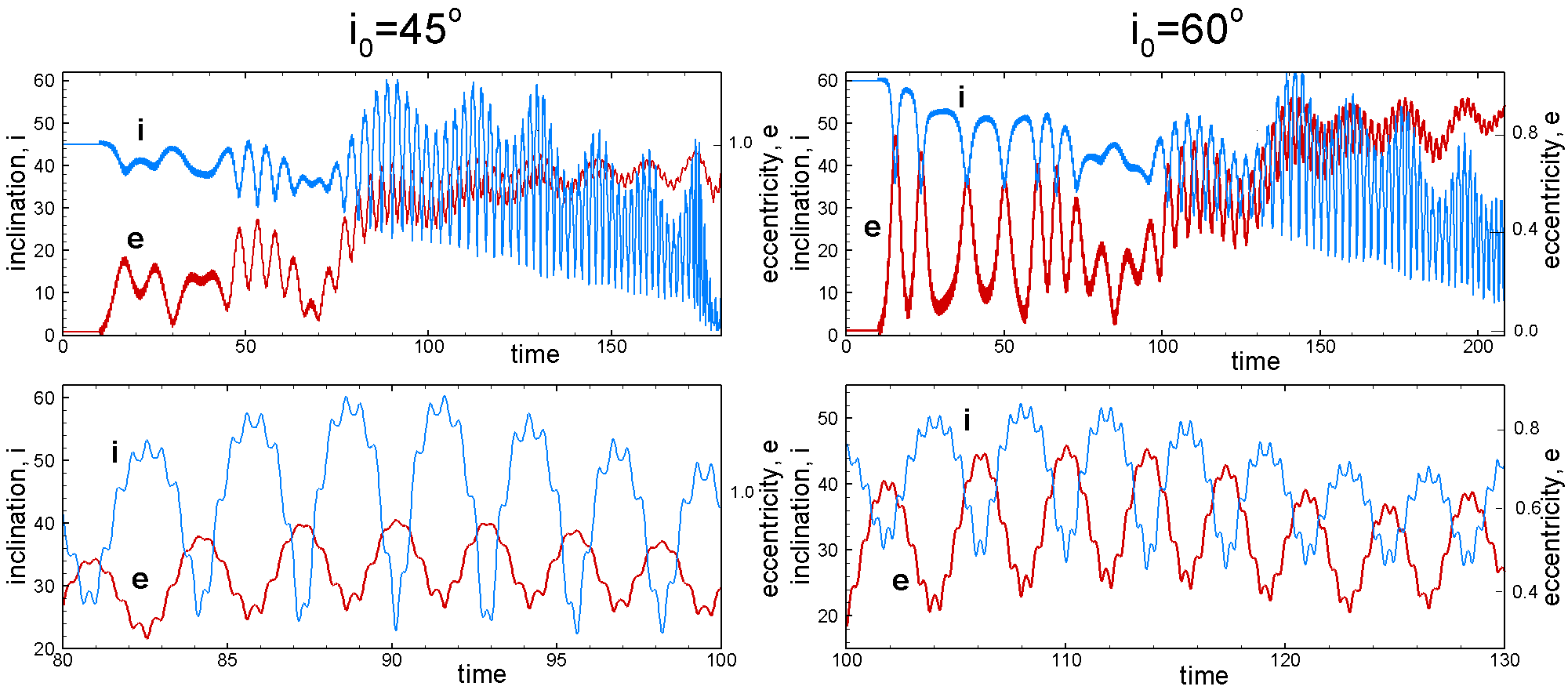} 
     \caption{\textit{Top panels:} Temporal evolution of the eccentricity, $e$ and inclination, $i$  in models \textit{n1.5i45} and \textit{n1.5i60}, where the eccentricity and inclination are shown at the same plot.
\textit{Bottom panels:} The same, but during shorter time intervals.
\label{fig:n1.5-ei-45-60}}
\end{figure*}

\section{Planets on inclined orbits: Kozai-Lidov effect} 
\label{sec:inclined}

Next, we placed a planet in an inclined orbit with different inclination angles $i_0$.  We observed that the eccentricity typically increases, and it also oscillates. The inclination angle also oscillates, but in antiphase with eccentricity. 
We suggest that we observe the Kozai-Lidov mechanism, where the disc acts as a massive object that perturbs the planet's orbit.  

\subsection{Planet-planet/star and planet-disc interaction}

Below, we briefly summarize the theory of the Kozai-Lidov mechanism in cases of the planet-planet/star interaction (e.g., \citealt{Lidov1962,Kozai1962,InnanenEtAl1997}), and planet-disc interaction \citep{TerquemAjmia2010}.

If a planet of mass $m_p$ located at an inclined orbit with semi-major axis $a_p$,
and interacts with a massive object (a planet or a star)  of mass $M_p$ located at the circular orbit of radius $R_p>>a_p$, then  the secular perturbation by the distant companion causes the eccentricity $e_p$ of the inner planet and the mutual inclination $i$ of two orbits  to oscillate in time in antiphase. 
In this situation, the component of the angular momentum 
of the inner orbit perpendicular to the orbital plane, $L_z$ is constant and proportional to
\begin{equation}
L_z\propto \sqrt{1-e_p^2} \cos{i}={\rm const}~.
\end{equation}
This equation shows that the decrease of the inclination angle $i$ leads to the increase of the eccentricity $e_p$, and vice versa.  
As a result, the eccentricity and inclination oscillate in the antiphase, and eccentricity can also be pumped to the orbit at the expense of inclination and vice versa.

\citet{TerquemAjmia2010} considered the interaction of the planet on the inclined orbit with the external remote disc and
noticed that the potential for the planet-planet interaction is similar to that for planet-disc interaction  (compare their formulae 4 and 5). 
They concluded that the Kozai-Lidov mechanism should also operate in the case of the remote discs.

The maximum value of the eccentricity which can be reached during this process is
\begin{equation}
e_{\rm max} =\bigg(1-\frac{5}{3} \cos^2 i_c\bigg)^{1/2} ~,
\label{eq:emax}
\end{equation}
and therefore the initial inclination $i_0$ should be larger than the critical value $i_c\approx 39^\circ$ which is determined from condition
$
\cos^2{i_c} = 3/5 ~.
$

The time $t_{\rm ev}$ to reach $e_{\rm max}$ starting from $e_0$ in both models 
is  \citep{InnanenEtAl1997}:
\begin{equation}
\frac{t_{\rm ev}}{\tau}=0.42 \bigg(\sin^2{i_0}-\frac{2}{5}\bigg)^{-1/2} \ln\bigg(\frac{e_{\rm max}}{e_0}\bigg)~,
\label{eq:tevol}
\end{equation}
\begin{equation}
\tau= K \bigg(\frac{R'}{a_p}\bigg)^3\bigg(\frac{M_*}{M'}\bigg) \frac{P}{2\pi} ~,
\label{eq:tau}
\end{equation}
where $P$ is the period of the planet's rotation. For planet-planet/star interaction: $M'=M_p$, $R'=R_p$ and $K=1$. For planet-disc interaction \citep{TerquemAjmia2010}:   $M'=M_d-$is the total mass of the disc, $R'=R_o$, and
\begin{equation}
K=\frac{(1+s)(1-\eta^{-s+2})}{(s-2)(1-\eta^{-s-1})}~,
\label{eq:K}
\end{equation}
where $\eta=R_{\rm in}/R_o$, $R_{\rm in}$ and $R_o-$are the inner and outer radii of the disc, $s-$is the power in the surface density distribution \citep{TerquemAjmia2010}.
If the eccentricity oscillates between $e_0$ and $e_{\rm max}$
then the period of oscillations is $P_{\rm osc}=2 t_{\rm ev}$.

\citet{TerquemAjmia2010} (and also  \citealt{TeyssandierEtAl2013}) performed numerical simulations using earlier developed N-body code \citep{PapaloizouTerquem2001} and confirmed  these theoretical results. In particular, they have shown that $t_{\rm ev}$
increases when $R_o$ increases,  and also $t_{\rm ev}$ increases when  the disc mass $M_d$ decreases which is in accord with eq. \ref{eq:tau}.
\citet{TeyssandierEtAl2013} confirmed the action of this Kozai-Lidov mechanism at various parameters of the model.

\begin{table*}
\begin{tabular}[]{ c  | c | c|c|c|c|c|c}
 \hline
             Parameter/model    &   $ n1.5i0$     & $n1.5i5$   &  $ n1.5i15$   & $n1.5i30$     & $n1.5i45$   & $n1.5i60$  & $n1.5i75$   \\
\hline          
  $i_0$                    &   $0^\circ$    &  $5^\circ$  &  $15^\circ$   & $30^\circ$   & $45^\circ$   & $60^\circ$    & $75^\circ$  \\   
    slope, $n$           &   $1.5$          &   $1.5$      &  $1.5$           &  $1.5$          &  1.5             &  1.5             &  1.5  \\
\hline 
 \hline
             Parameter/model    &   $ n1.8i0$     & $n1.8i5$   &  $ n1.8i15$   & $n1.8i30$     & $n1.8i45$   & $n1.8i60$  & $n1.8i75$   \\
\hline          
  $i_0$                    &   $0^\circ$    &  $5^\circ$  &  $15^\circ$   & $30^\circ$   & $45^\circ$    & $60^\circ$    & $75^\circ$  \\   
    slope, $n$           &   $1.8$          &   $1.8$      &  $1.8$           &  $1.8$          &  1.8             &  1.8            &  1.8         \\
\hline 
\end{tabular}
\caption{Simulations were performed at a variety of the initial inclination angles of the planet's orbit, $i_0$, and two values of the density slope in the disc, $n=1.5$ and $n=1.8$.
 Names of models are constructed using the values of $n$ and $i_0$. 
  \label{tab:ref-model-name}}
\end{table*}

\begin{figure*}[h]
     \centering
     \includegraphics[width=0.8\textwidth]{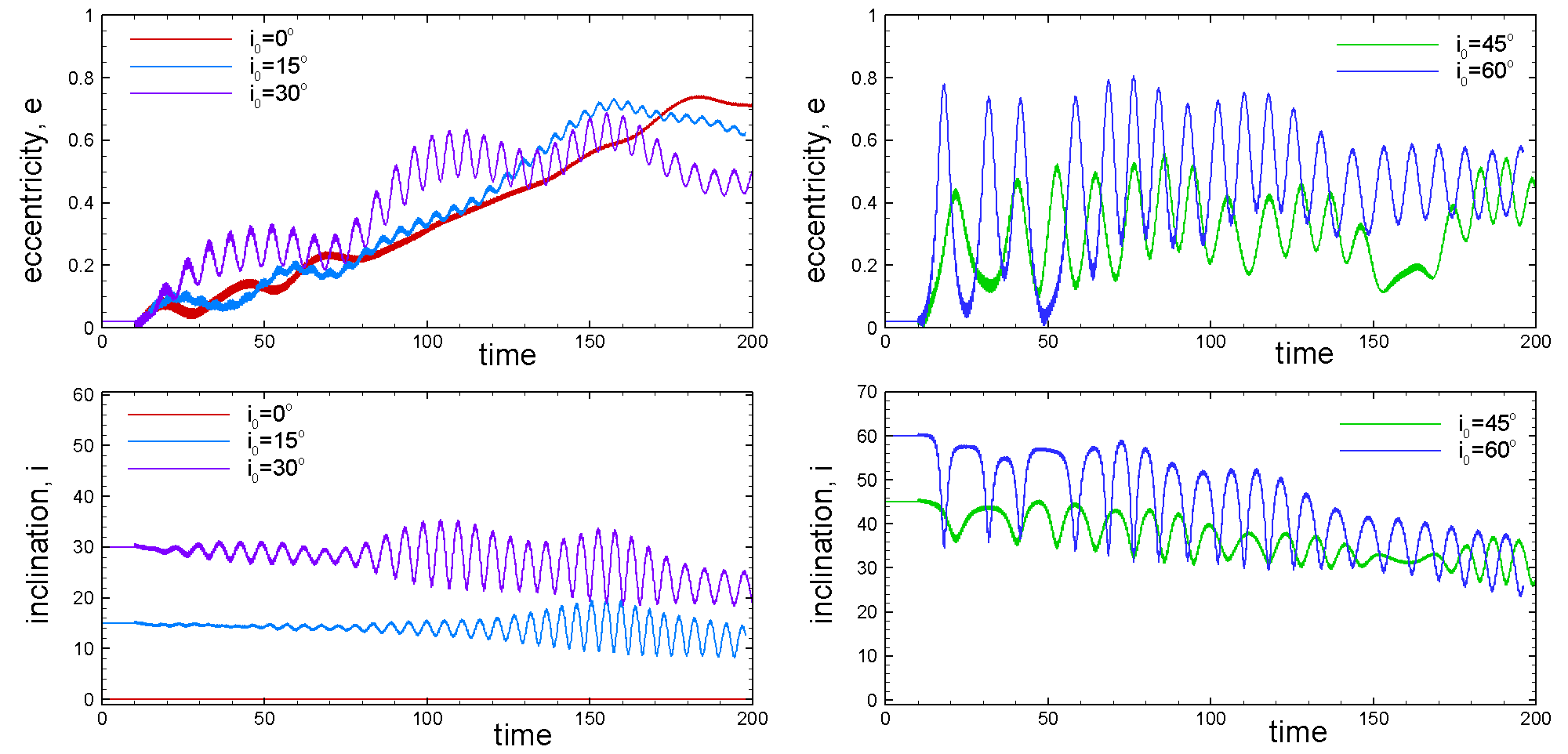} 
     \caption{The same as in Fig. \ref{fig:n1.5-ei-4} but for reference models with $n=1.8$ and different $i_0$.
\label{fig:n1.8-ei-4}}
\end{figure*}

 \subsection{3D simulations of planets on inclined orbits}

We put a planet on inclined orbits with different inclination angles $i_0$ and two types of discs with 
 $n=1.5$ and $1.8$.

\subsubsection{Disc-planet interaction in model $n1.5i45$}

As an example, we take one of the reference models, $n1.5i45$, and show results for the disc-planet interaction in detail.

Top panels of Fig. \ref{fig:n1.5-45-all}  and the 3rd row from the top 
show that ELR waves form in the inner disc both in the surface density (top row) and the equatorial density distribution  (3rd row). The $m=2$ modes are clearly observed. The 2nd row shows the density distribution in the disc. The 4th row shows the equatorial density distribution, which decreases rapidly with radius. The bottom left panels show that the eccentricity oscillates but increases on average up to $e\sim 0.7-0.9$ after $t\approx 70$. The inclination strongly oscillates and decreases on average. The bottom right panels show the variation of the planet's coordinates (top)  and the $rz-$slice of the density distribution (bottom). One can see that in equatorial $x-$ and $y-$ coordinates, the planet reaches the inner disc radius of $r\sim 1-1.2$, which leads to a variation of its eccentricity around a large value of $\sim 0.8$. There are also large-scale variations of the $x-$ and $y-$ coordinates due to the precession of the orbit. The $z-$component decreases with time. The bottom right panel shows that the disc is not symmetric about the equatorial plane due to the disc-planet interaction.

\subsubsection{Models with $n=1.5$ and different $i_0$}

We took our reference model with $n=1.5$ but placed a planet in orbits with different inclination angles:
$i_0=5^\circ, 15^\circ,  30^\circ,  45^\circ, 60^\circ,75^\circ$. 

 We observed that in models with relatively small inclination angles,
$i_0=5^\circ - 30^\circ$, ELR resonances were excited in the inner disc,  which are similar to those in the model with $i_0=0^\circ$. The eccentricity evolution and the growth rates are similar to those in the model with $i_0=0^\circ$ (see the top
 left panel of Fig. \ref{fig:n1.5-ei-4}). The bottom left panel shows that the inclination angle decreased on average but strongly oscillates after time $t>50-70$. 
At inclination angle $i_0=30^\circ$, the eccentricity increases faster than in models with smaller $i_0$. 

At larger inclination angles, $i_0=45^\circ$ and $i_0=60^\circ$, the eccentricity initially strongly oscillates and reaches $e\approx 0.6$. Later, it increases to higher values 
of  $e\approx 0.7-0.9$. The inclination of the orbit decreases on average and strongly oscillates (see right panels in Fig. \ref{fig:n1.5-ei-4}). 

The top panels of Fig. \ref{fig:n1.5-ei-45-60} show variation of eccentricity and inclination which were placed side by side to the same  panels. 
The bottom panels show a part of the simulation time with a higher temporal resolution. One can see that the inclination and eccentricity oscillate in the antiphase, as predicted by the Kozai-Lidov mechanism. We observed such antiphase oscillations in all simulation runs with inclined orbits. We think we observe the eccentricity growth and its oscillations due to the Kozai-Lidov mechanism. 


\begin{figure*}
     \centering
     \includegraphics[width=0.8\textwidth]{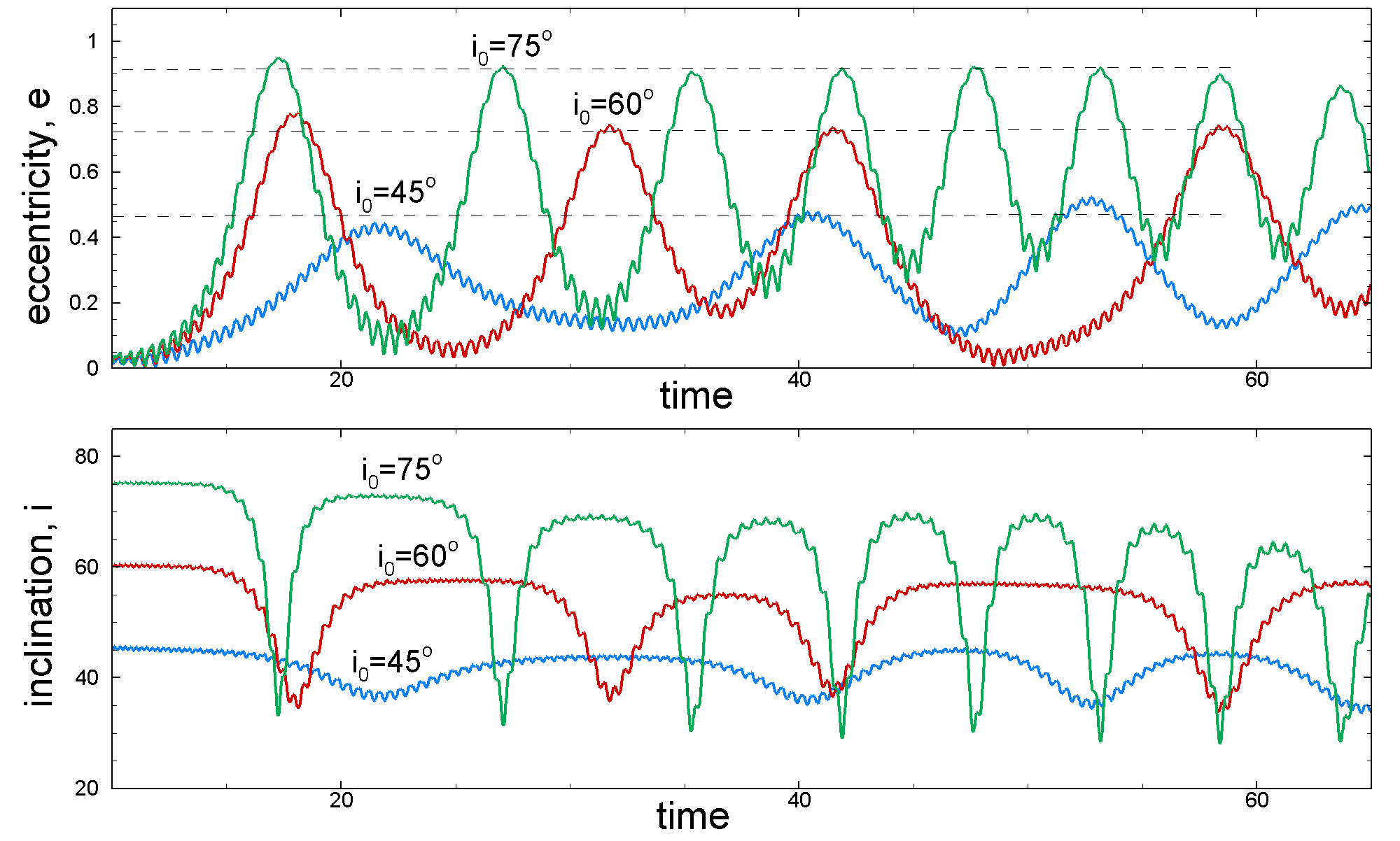} 
     \caption{\textit{Top panel:} Eccentricity evolution in the model with $n=1.8$ at different initial values of the inclination angle of the orbit: $i_0=45^\circ, 60^\circ, 75^\circ$. Dashed horizontal lines show an approximate value of the eccentricity amplitude for each model. \textit{Bottom panel:} Same, but for the inclination angle.
\label{fig:Koz-i-n1.8-2}}
\end{figure*}

\begin{figure}
     \centering
     \includegraphics[width=0.45\textwidth]{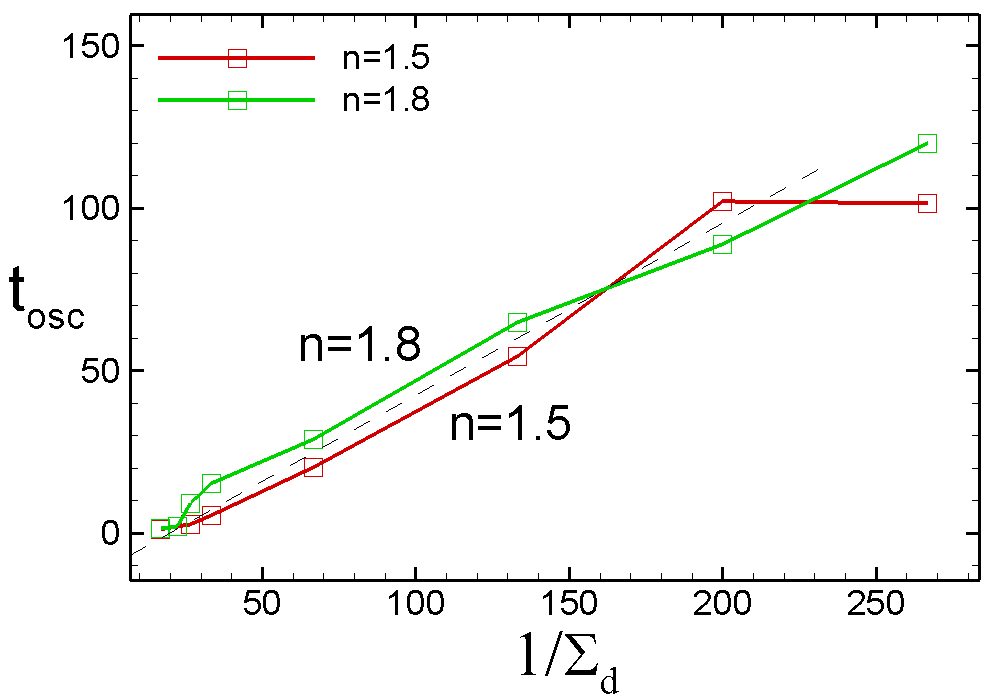} 
     \caption{Period of Kozai-Lidov oscillations obtained in simulations of discs with different initial surface densities in models $n1.5i45$ and $n1.8i45$. The dashed line shows the dependence taken for analytical estimates and projections in Eqs. \ref{eq:tau} and \ref{eq:tau-main}.
\label{fig:Koz-tau-sig-45}}
\end{figure}

\subsubsection{Models with $n=1.8$ and different $i_0$}

We repeat the above simulations using a reference model with a steeper density distribution, $n=1.8$.
The top left panel of Fig. \ref{fig:n1.8-ei-4} shows that in models with relatively small inclination angles,  $i_0=15^\circ$ 
and $i_0=30^\circ$, the eccentricity increases similar to that in the model with zero inclination ($n1.8i0$). 
However, the eccentricity increases $\sim 1.8$ times slower compared with models where $n=1.5$. 
The bottom left panel 
shows that the inclination oscillates and decreases on average.

The right panels of the same figure show the eccentricity and inclination in models with high inclination angles. 
One can see 
that the eccentricity strongly oscillates and reaches values of $e\approx 0.55$ in the model with $i_0=45^\circ$ and 
$e\approx 0.8$ in the model with $i_0=60^\circ$. The inclination angle also strongly oscillates and decreases on average. 
The time scale of oscillations is $\sim 3$ times longer in models with $n=1.8$ compared with models $n=1.5$. 
We suggest that this is because at $n=1.8$, the disc has a lower mass compared with $n=1.5$ models. 
  Eq. \ref{eq:tau} from the theory shows that the time scale of oscillations is inversely proportional to the mass of the disc, $M_d$.
 The mass of the disc in the model $n1.8i45$ is approximately 1.8 times smaller than that in the model $n1.5i45$. Comparisons show a correct tendency towards longer time scales.



\subsubsection{Dependence of the maximum eccentricity $e_{\rm max}$ on $i_0$} 

According to the theory, the maximum value of the eccentricity should increase with inclination of the orbit (see Eq. \ref{eq:emax}).
From this equation it follows that for inclinations  $i_0=45^\circ, 60^\circ$ and $75^\circ$, the maximum eccentricities are $e_{\rm max}\approx 0.41, 0.76, 0.94$, respectively.

We took a model with $n=1.8$ and compared maximum eccentricity values in models with $i_0=45^\circ, 60^\circ$ and $75^\circ$. We chose early moments before other processes started to influence the eccentricity growth. Top panel of 
Fig. \ref{fig:Koz-i-n1.8-2} shows that $e_{\rm max}$ is larger in models with larger initial inclination.  From the plot  (see dashed horizontal lines in the plot), 
we obtain:: $e_{\rm max}\approx 0.46, 0.73,0.92$ for models with  $i_0=45^\circ, 60^\circ$ and $75^\circ$, respectively. 
These values are very close to those predicted by the theory. 
We suggest that our model is close to the theoretical model by \citet{TerquemAjmia2010} because at $n=1.8 $, most of the mass
is in the outer regions of the disc, which is close to the theoretical model, where the disc is located far away from the planet.
 The bottom panel of Fig.  \ref{fig:Koz-i-n1.8-2} shows that the amplitude of inclination also  increases with $i_0$.

\subsubsection{Time scale of eccentricity growth in oscillations}

In this experiment, we fix the disc radii and structure but change the reference surface density $\Sigma_d$. This way, we change the mass of the disc. We observed that the time scale of oscillations increases when $\Sigma_d$ decreases. Fig. \ref{fig:Koz-tau-sig-45}  shows the dependence of the period of oscillations $t_{\rm osc}$ on  $1/\Sigma_d$ is similar in models with $n=1.5$ and $n=1.8$.  The dependence is  approximately linear for  $1/Sigma_d \gtrsim 100$. From the plot, we derive an approximate dependence:
\begin{equation}
t_{\rm osc}\approx 0.53(\Sigma_d^{-1}-18)~.
\label{eq:tosc}
\end{equation}
At small values of $\Sigma_d$, we obtain  $t_{\rm osc}\approx 0.53 \Sigma_d^{-1}$.
In our models, the mass of the disc $M_d\sim \Sigma_d$ and therefore, $t_{\rm osc}\sim 1/M_d$, as predicted in theoretical models (see Eq. \ref{eq:tau}).

 Equation  \ref{eq:tosc} is in dimensionless units.
For practical applications, we convert this equation to dimensional units using the projected value of the characteristic disc mass of  $\Sigma_d=10^{-4}$.  We take into account that we measure time in rotational periods at $r=r_{\rm cav}$  and take $r_{\rm cav}=10$ AU as a reference scale  (see Tab. \ref{tab:units}), and obtain :
\begin{equation}
T_{\rm osc}\approx 3.18\times 10^5 {\rm yr} \bigg(\frac{r_{\rm cav}}{10AU}\bigg)^{3/2} \bigg(\frac{\Sigma_d}{10^{-4}}\bigg)^{-1}  ~.
\label{eq:tau-main}
\end{equation}
This time scale is comparable with the time scale of eccentricity growth due to ELRs (see Eq. \ref{eq:t-ecc}).

\begin{figure*}[b]
\centering
\includegraphics[width=0.4\textwidth]{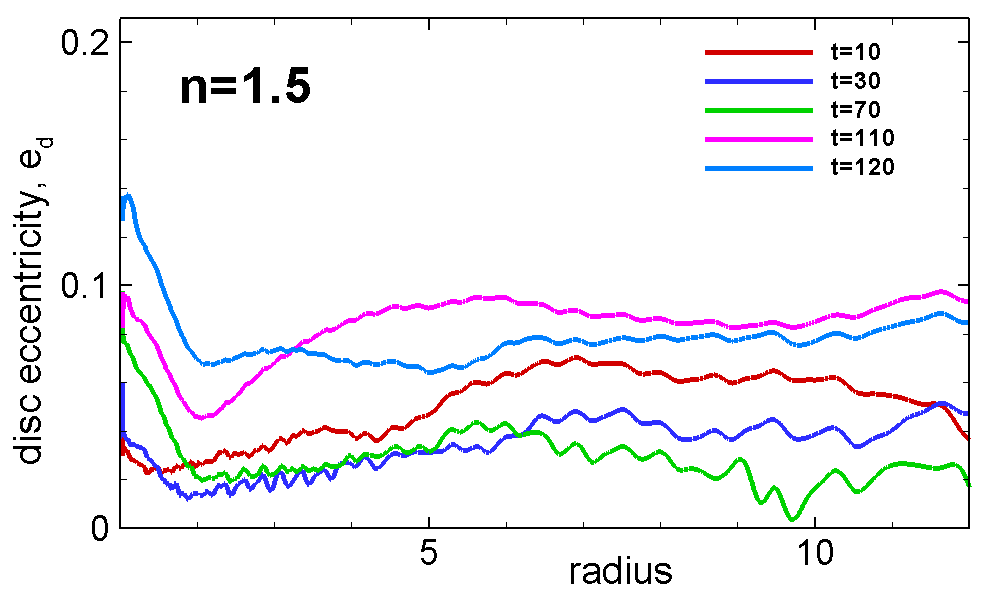} 
\includegraphics[width=0.4\textwidth]{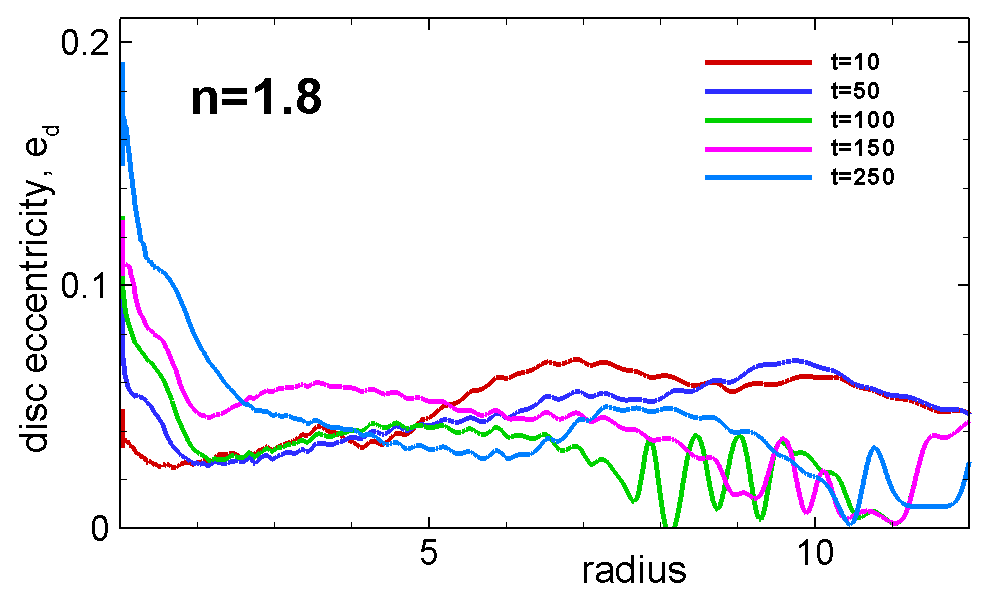} 
\caption{Radial distribution of the disc eccentricity at different moments of time 
 obtained in models  $n1.5i0$ (left panel) and $n1.8i0$ (right panel). 
\label{fig:ecc-disc}}
\end{figure*}

We also compared the time scale of eccentricity growth $t_{\rm ev}$ given by the theory (see equations \ref{eq:tevol} - \ref{eq:K})  with that obtained in our simulations.  As an example, we use a model $n1.5i60$, which shows high-amplitude oscillations, and consider the 3-rd peak in the eccentricity curve shown in the top right panel of Fig. \ref{fig:n1.5-ei-45-60} (we take one of the early moments when the disc is approximately homogeneous and the density wave did not form yet). 
 In the 3-rd peak $e_0\approx 0.1$,
$e_{\rm max}\approx 0.6$. We take $a_p=0.6$ and $R_d=13$ (the radius of exponential cut) and obtain $\eta\approx 0.077$.   We also obtain $K\approx 0.041$, $t_{\rm ev}/\tau\approx 1.27$. We calculate the dimensionless mass of the disc as $M_d\approx 15.8$ and obtain the final value of time in our dimensionless units as $t_{\rm ev}\approx 15.7$.  We compare this value with the time of eccentricity growth in the 3rd peak obtained from the figure, which is $t_{\rm sim}\approx 9$. One can see that the difference is in the factor of 1.7, which is in reasonable agreement with the theory. 


\begin{table*}
\begin{tabular}[]{ c  | c | c | c | c| c |c|c|c|c }
\hline 
 Reference desity  $\rho_d$                             &   0.8      & 0.6       &  0.4       &  0.3       &  0.2      & 0.133  &  0.1        & 0.068      & 0.05        \\   
Reference surface density   $\Sigma_d$    &  0.06     &  0.045   & 0.03      & 0.0225   & 0.015     & 0.01         & 0.0075    &   0.0051          & 0.00375         \\
\hline
\end{tabular}
\caption{Densities $\rho_d$  and corresponding surface densities   $\Sigma_d$ used to derive the dependence of the oscillation time $t_{\rm osc}$ from the inner density of the disc (see  Fig. \ref{fig:Koz-tau-sig-45}).
 \label{tab:q_d}}
\end{table*}

\begin{table}
\begin{tabular}[h]{ cc | cccc}
\hline          
                 m & res & $r_{\rm res}/a_p$  &  $\cal A$    &  $\cal B$ &  ${\cal A}/{\cal B}$       \\
\hline  
                 2 & 1:3 & 2.080                 & 0.607       & 1.849      &   0.328      \\ 
                 3 & 2:4 & 1.587                 & 5.201       & 3.594      &   1.447     \\
                 4 & 3:5 & 1.406                 & 7.362       & 5.604      &   1.314     \\
\hline 
\end{tabular}
\caption{The mode number $m$, type of resonance, ${\rm res}$, resonant radii $r_{\rm res}/a_p$, coefficients  $\cal A$    and  $\cal B$ for ELRs and their ratios.
  \label{tab:resonances}}
\end{table}

\begin{figure*}
\centering
\includegraphics[width=0.8\textwidth]{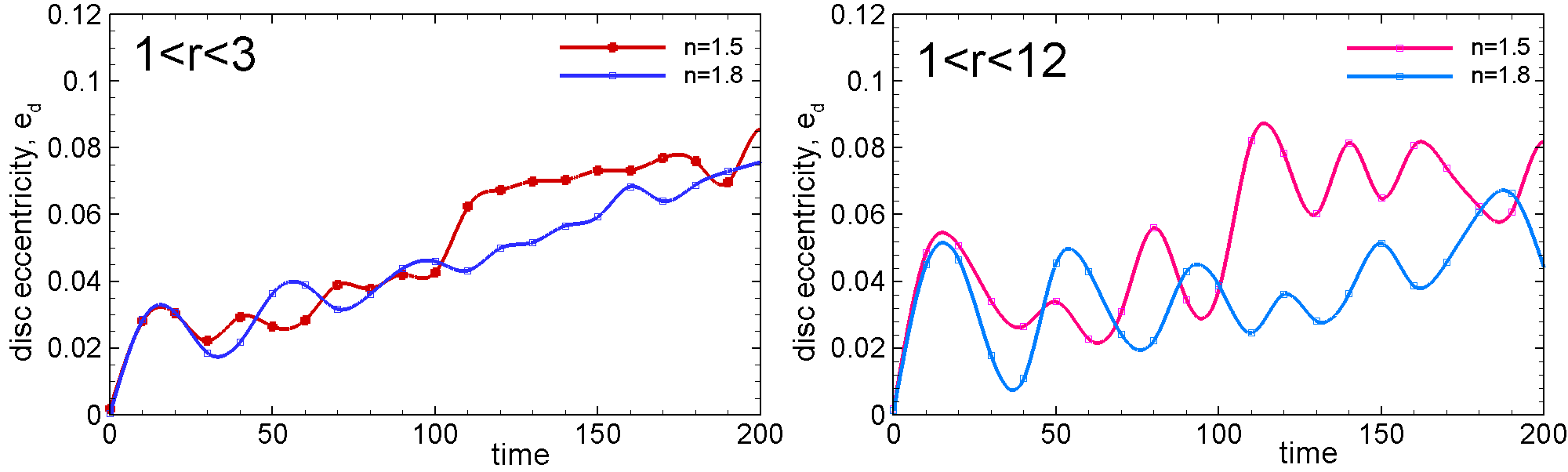} 
\caption{\textit{Left panel:} Temporal variation of the averaged eccentricity of the disc $\bar{e_d}$ in models   $n1.5i0$ and $n1.8i0$ calculated for the inner part 
of the disc, 
in the interval of radii $1<r<3$. \textit{Right panel:} Same, but for the disc eccentricity taken in the interval of radii  $1<r<12$. 
\label{fig:ecc-disc-time}}
\end{figure*}

\begin{figure*}
\centering
\includegraphics[width=0.49\textwidth]{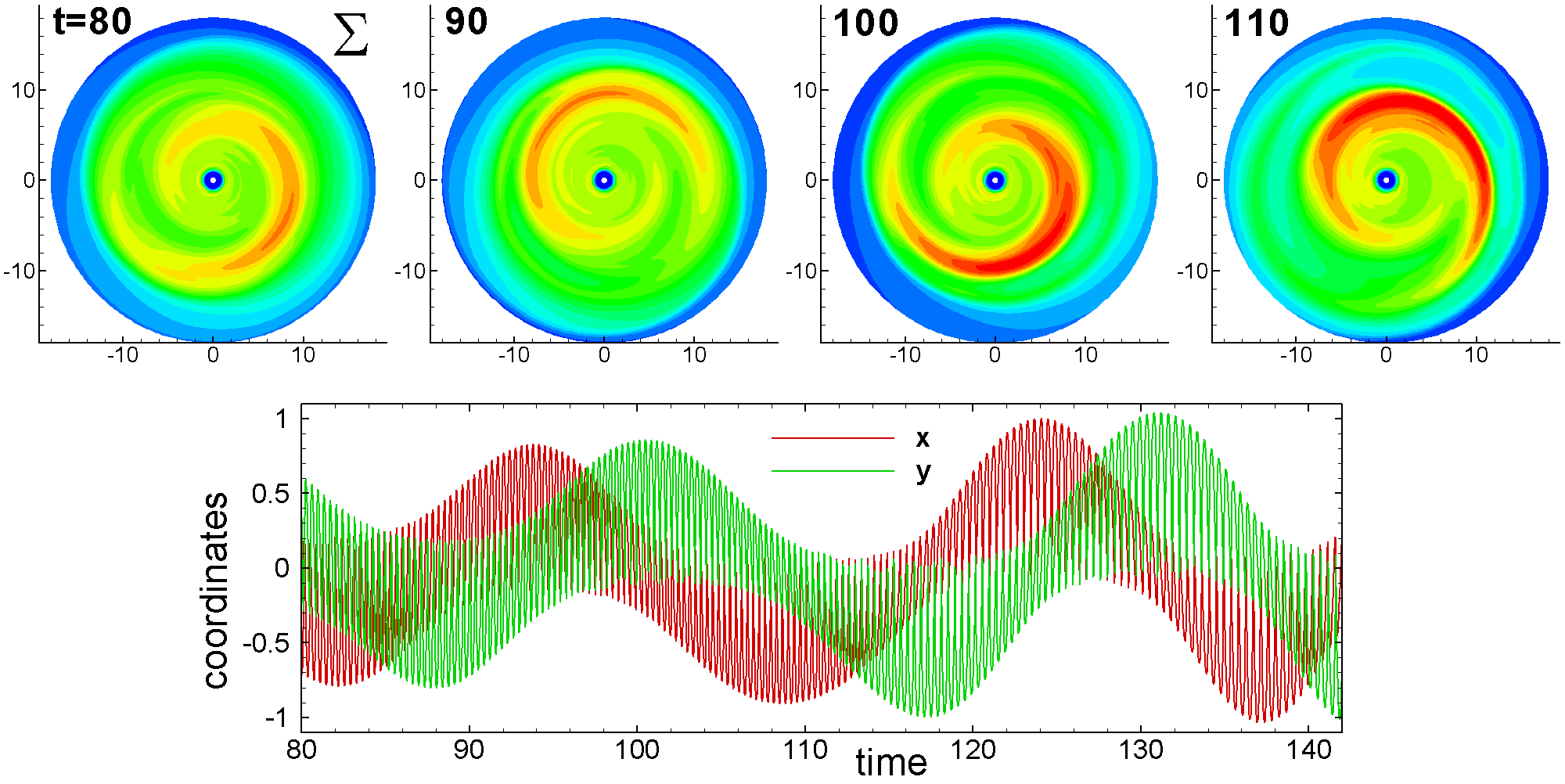} 
\includegraphics[width=0.49\textwidth]{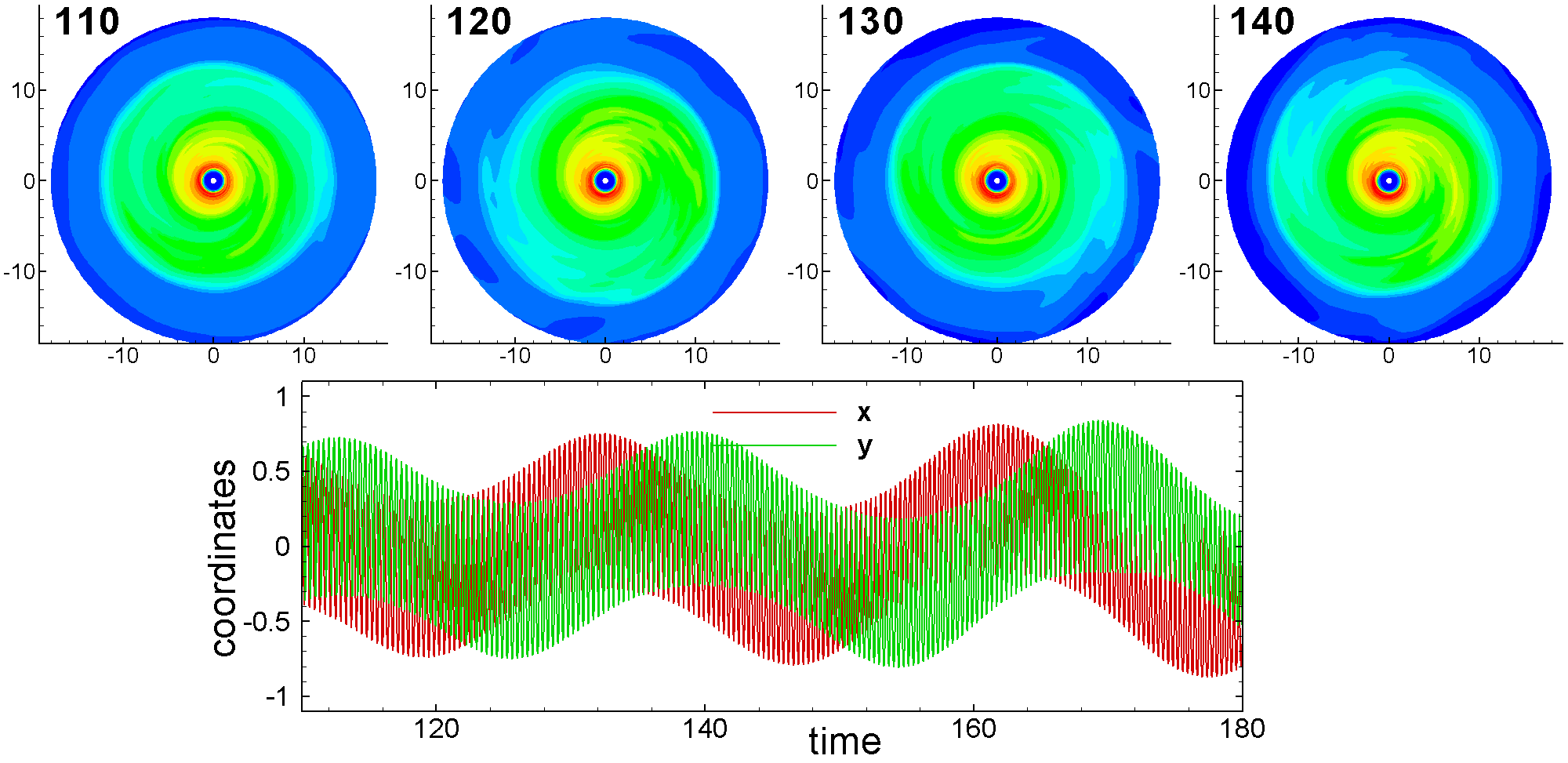} 
\caption{{\textit{Top panels:} Density distribution in the disc during the sequence of times shows the precession of the disc in models  $n1.5i0$ (left four panels) and $n1.8i0$ (right four panels). 
\textit{Bottom panels:} Variation of $xy-$coordinates of the planet in the equatorial plane shows the precession of the planet. }
\label{fig:n1.5-1.8-prec}}
\end{figure*}

\section{Eccentricity of the disc and precession} \label{sec:disc-ecc}

A planet on the eccentric orbit excites eccentricity in the disc (e.g., \citealt{Ogilvie2007}).
The linear analysis (performed for small eccentricities of the planet and the disc) shows that the eccentricities of the planet and the disc 
are coupled through resonant interactions (e.g., \citealt{Ogilvie2007,TeyssandierOgilvie2016}).  
To compare our results with theory, we 
use formulae for the temporal evolution of complex  eccentricities  from 
\citet{TeyssandierOgilvie2016} (see their Eqs. 14 and 15). 
A single 
ELR contributes to the evolution  of eccentricities of the planet and the disc (in the vicinity of the resonance)  in the following way:
 \begin{equation}
M_p a_p^2\Omega_p\bigg(\frac{\partial E_p}{\partial t}\bigg)_{\rm ELR}=\frac{GM_p^2}{M_\star}E_p {\cal B}^2\bigg(1 - \frac{{\cal A}E_d}{{\cal B} E_p}\bigg)\int{\Sigma }
F 2\pi r dr
\label{eq:dedt-p}
\end{equation}
 \begin{equation}
\Sigma r^2\Omega\bigg(\frac{\partial E_d}{\partial t}\bigg)_{\rm ELR}=- \frac{GM_p^2}{M_\star}{\Sigma{\cal A}{\cal B}E_p \bigg(1-\frac{{\cal A} E_d}{{\cal B}E_p}\bigg)} F  ,
\label{eq:dedt-d}
\end{equation}
where
$E_p=e_p e^{i\bar{w_p}}$ and  $E_d=e_d e^{i\bar{w_d}}$, 
$e_p=|E_p|$, $e_d=|E_d|$; 
$\bar w_p$ and $\bar w_d$ are the arguments of the pericentre of the planet's and disc's semimajor axes, respectively. 
Here
$
F=w_L^{-1}\Delta [(r-r_{\rm res})/{w_L}-1]
$
is a function of resonant radius $r_{\rm res}$, resonant width $w_L$ and dimensionless function $\Delta$.
 Values  $w_L$ and function $\Delta$ describe the radial profile of the ELR  resonance.  

In our simulations, we observe formation of ELR waves with mode numbers $m=2, 3$ and sometimes $m=4$. Tab. \ref{tab:resonances}
 shows the values of coefficients  $\cal A$ and $\cal B$ for these resonaces and the resonant radii $r_{\rm res}$ 
 (see an extended version of the table in \citealt{TeyssandierOgilvie2016}).

In the above sections, we calculated the evolution of the planet's eccentricity. Below, we study the disc eccentricity and precession of the planet and disc.

\subsection{Disc eccentricity}

We calculate the distribution of the disc eccentricity with radius using an approach based on the angular momentum deficit  (hereafter AMD) $A_d(r)$  (e.g., \citealt{RagusaEtAl2018}). 
The angular momentum deficit of the ring is
$
A_d(r) =J_{\rm circ}(r) - J_d(r)~,
$
where 
$$
J_{\rm circ}(r) = \int \Sigma \sqrt{GM a}~d\phi~,~~~~a = - \frac{GM}{2E}~,~~~~E = - \frac{GM}{r} + \frac{v^2}{2}
$$
is the circular angular momentum of the ring in the disc located at radius r, and $
J_d (r)= \int \Sigma~r v_\phi d\phi .
$
is the real angular momentum of the ring at the radius $r$.
The eccentricity of the ring is
\begin{equation}
e_d(r) = \sqrt {\frac{2~A_d(r)}{J_{\rm circ}(r)} } .
\end{equation}

Our disc has a finite thickness and, therefore, is not precisely Keplerian due to the pressure component. 
In the calculation of the disc eccentricity, we subtracted this background eccentricity
 (see also \citealt{RagusaEtAl2024}).

Fig. \ref{fig:ecc-disc} shows the distribution of $e_d(r)$ at different moments in time in models $n1.5i0$ and $n1.8i0$ (see left and right panels, respectively).   In both models, the disc eccentricity is larger in the inner disc (in the region of ELRs).
Eccentricity in the inner disc increases with time in both models. Eccentricity in the rest of the disc increases most of the time in the model $n1.5i0$ and reaches $e_d\approx 0.1$ on average at $t= 110$. 
In the model $n1.8i0$ the disc eccentricity varies and is $e_d\approx 0.05-0.06$ on average. 

Next, we  calculate the evolution of eccentricity with time.  For that, we take the average eccentricity value in some radii interval. The left panel of  Fig. \ref{fig:ecc-disc-time} shows the temporal evolution of eccentricity where we took the averaged value in the interval of radii   $1<r<3$ (where ELR resonances are located). One can see that the inner disc
eccentricity gradually increases in both models.  The right panel shows the temporal variation of eccentricities averaged at $1<r<12$. One can see that the eccentricity of the whole disc varies quasi-periodically,  with a quasi-period of $30-40$ in model $n1.5i0$ and a slightly longer quasi-period in model $n1.8i0$.

Now, we can compare simulation results with theoretical expectations.
Here, we neglect the precession 
 and take the absolute values, $e_p=|E_p|$  and $e_d=|E_d|$ (like we did in R23).
According to the theory (see Eq.  \ref{eq:dedt-p}), the planet's eccentricity will increase if   the value in the brackets
$(1 - {{\cal A}e_d}/{{\cal B} e_p})>0$, that is if $e_p > {\cal A}/{\cal B} e_d$. As an example, we take a model 
 $n1.5i0$ and some moment in time $t=50$. From the left bottom panel of Fig.  \ref{fig:n1.5-all}, we obtain the planet eccentricity  $e_p\approx 0.25$,  and from the left panel of  Fig. \ref{fig:ecc-disc-time} the inner disc eccentricity:   $e_d\approx 0.025$.  Simulations show that at this time, ELR $1:3$ resonance dominates ($m=2$). Taking the value of ${\cal A}/{\cal B}$ for 1:3 resonance  from Tab. \ref{tab:resonances} and $e_d=0.025$, we obtain that condition for planet eccentricity growth  becomes $e_p>
0.008$. 
This condition is satisfied, and the planet's eccentricity will grow. Comparisons at other times show a similar result, and therefore, in our model, conditions are always favorable for the planet's eccentricity growth.

Similar estimates for the disc eccentricity growth (see Eq. \ref{eq:dedt-d}) show that the disc eccentricity cannot grow at any values of $e_d$ and $e_p$ obtained in simulations.   We suggest that relatively small eccentricities of the disc observed in our simulations may be connected with this theoretical prediction.

\subsection{Precession}

Simulations show that the planet's orbit precesses counterclockwise. The bottom left panels of 
Fig. \ref{fig:n1.5-all} and Fig. \ref{fig:n1.8-all} 
show coordinates $x$ and $y$ of the planet's orbit in the equatorial plane. The waves in the curve reflect the precession of the planet.  The amplitude of waves increases due to the increase of eccentricity. Period of planet precession is $T_{\rm prec}\approx 25-30$ in model $n1.5i0$ and $T_{\rm prec}\approx 30-35$ in model $n1.8i0$. Test simulations of model $n1.5i0$ at a twice as low 
and high inner surface density of the disc $\Sigma_d=  0.015$ and $\Sigma_d=  0.06$ have shown that the period of precession is larger in models with a lower density of the disc.
 The disc also precesses. The precession can be tracked using the orientation of the 
density wave seen in the surface density distributions
(see Fig. \ref{fig:n1.5-all})\footnote{The density wave tracks the precession of the disc only approximately. \citet{TeyssandierOgilvie2016} note that eccentricity excited in the inner parts of the disc propagates out in the form of a one-armed density wave. 
It is probable that
 the density wave results from the fact that 
initially, the rings of matter in the disc have different precession rates and  different lines of percenters, which are more aligned in the inner parts of the disc where processes are faster
(see an illustration of this effect in Fig. 1 of \citealt{RagusaEtAl2024}).}.

Fig. \ref{fig:n1.5-1.8-prec} shows an episode of precession in models $n1.5i0$ and $n1.8i0$ in greater detail. The top left panels show the time sequence of the surface density distribution during an interval of time $\Delta t=80-110$ in model $n1.5i0$.  It shows that the density wave in the disc  precesses counterclockwise with a period of  $T_{\rm prec}\approx 25-30$. The bottom left panel shows
that a planet precesses approximately with the same period. 

Right top panels of Fig. \ref{fig:n1.5-1.8-prec} show the same but for the model $n1.8i0$. The top panels show that the spiral wave is only slightly visible. We suggest that in this model, where the surface density decreases with radius, lines of pericentres become aligned more rapidly than in the previous model of homogeneous disc.  The bottom panel shows that the planet precesses with a period of $T_{\rm prec}\approx 30-35$.

According to Eqs. \ref{eq:dedt-p} and \ref{eq:dedt-d}, 
the precession of the planet and disc can influence the rate of eccentricity growth \citep{TeyssandierOgilvie2016}.  
From simulations, we see that the planet and disc precess couterclockwise approximately at the same rate.  
If they precess with precisely the same rate and have the same phase, then  ${\bar w}_p-{\bar w}_d=0$,  and $e^{i({\bar w}_p-{\bar w}_d)}=1$. 
 In the opposite situation, if they precess in antiphase,  $\omega_p-\omega_d=\pi$, and  $e^{i(\omega_d-\omega_p)}=-1$. 
In both cases, the value in brackets in the right-hand side of Eq.  \ref{eq:dedt-p} is positive and planet's eccentricity will grow.

\section{Conclusions} \label{sec:conclusions} 

We have investigated the evolution of the eccentricity  of
massive planets located  inside cavities of protoplanetary discs.   
 The main conclusions  are the following:

\smallskip

\noindent \textbf{1.  In models with aligned orbits  ($i_0=0$):}

\begin{itemize}

\item   The eccentricity increases up to high values of $e\sim 0.7-0.9$  due to the ELR resonances excited in the inner disc. Resonances with modes $m=2$ and $m=3$ dominate. The eccentricity increases any time when ELR waves are excited in the disc. This process is similar to that observed in 2D simulations of R23. 

\item   The characteristic time of eccentricity growth increases in models with smaller planet mass due to smaller torque acting on the disc.  The amplitude of ELR waves is smaller in models with smaller planet mass.

\item At higher viscosity in the disc, the ELR density waves become  smeared, and eccentricity growth decreases. 

\item  The grid resolution is an essential factor. At a low grid resolution, the number of grids could not be sufficient to resolve ELR waves. In addition, the amplitude of density waves decreases due to the numerical diffusivity.

\item The disc eccentricity slowly increases with time, with the largest eccentricity at the inner disc. It increases with the growth of the planet's eccentricity. Planet-disc interaction leads to the precession of the planet's orbit. 
The density waves or other inhomogeneities in the disc precess with comparable period.
Disc eccentricity and its influence on the planet's orbit should be further studied in models with lower disc density and longer simulation runs.

\end{itemize}

\smallskip

\noindent \textbf{2. In models with inclined orbits ($i_0\neq 0$):}


\begin{itemize}

\item At relatively small inclination angles, $i_0\lesssim 30^\circ$, the eccentricity increases up to $e\sim 0.7-0.9$  due to the ELRs,   like in models with $i_0=0$. The Kozai-Lidov oscillations of small amplitude are observed. The orbital inclination decreases on average.

\item At large inclination angles, $i_0=45^\circ$,   $i_0=60^\circ$, and $i_0=75^\circ$ eccentricity and inclination strongly oscillate in the antiphase, like in the original Kozai-Lidov mechanism. 

\item The amplitude of oscillations increases 
when $i_0$ increases and reaches $e\approx 0.9$ in the case of $i_0=75^\circ$. 
The time scale of eccentricity growth increases when the characteristic disc's mass decreases.  

\item Eccentricity may also increase on average due to ELRs. 

\end{itemize}

The above simulations show good potential for explaining the eccentricity of exoplanets, including very high eccentricities. 
However, the final eccentricity at the time of disc dispersal can be different, and it depends on a number of factors. 
 One of the important factors is 
 the size of the disc-cavity boundary. At relatively small sizes, say, at  $r_{\rm cav}\lesssim10$AU, the eccentricity increases 
rapidly, and the planet may enter the inner disc and lose eccentricity due to the local corotation torque. Later, if the disc will move away,  the eccentricity may increase again.   Therefore, several episodes of eccentricity growth and decay may occur. In the case of planets on inclined orbits, several Kozai-Lidov cycles of eccentricity oscillation are expected. On the other hand, for cavities located at much larger distances, the eccentricity increases slowly, and only a part of the eccentricity growth cycle is expected.

 Our simulations were performed at a high density (and mass) of the disc, which provided  higher torques between the planet and the disc and helped to decrease computing time. We scale simulations to lower densities and longer time scales using the theoretical prediction that
the  eccentricity growth rate is inversely proportional to the density
 (e.g., \citealt{GoldreichTremaine1978}) and results of 2D simulations of R23, which confirmed this dependence. Future 3D simulations should be done at the lower density in the disc and also in discs with a steeper density distribution.

\section*{Acknowledgments}
The authors thank the anonymous referee for valuable recommendations. 
Resources supporting this work were provided by the NASA High-End
Computing (HEC) Program through the NASA Advanced Supercomputing
(NAS) Division at Ames Research Center and the NASA Center for
Computational Sciences (NCCS) at Goddard Space Flight Center. MMR and RVEL were supported
in part by the NSF grant AST-2009820. C. Espaillat was supported in part by NASA ADAP 80NSSC20K0451.

\section{Data Availability}

The data underlying this article will be shared on reasonable request to the corresponding author (MMR).

\bibliographystyle{mn2e}

\end{document}